\documentclass{article}

\usepackage[margin=0.7in]{geometry}
\usepackage[english]{babel}
\usepackage[utf8]{inputenc}
\usepackage{graphicx}
\usepackage{graphics}
\usepackage{fancyhdr}
\usepackage{amsmath} \numberwithin{equation}{section}
\usepackage{amsthm}
\usepackage{mathrsfs}
\usepackage{mathtools}
\usepackage{slashed}
\usepackage{amssymb}
\usepackage{braket}
\usepackage{bm}
\usepackage{wrapfig}
\usepackage{adjustbox}
\usepackage{multicol}
\usepackage{float}
\usepackage{url}
\usepackage{xcolor}
\usepackage{pagecolor}
\usepackage{subcaption}
\usepackage{tensor}
\usepackage{multirow}
\usepackage{xcolor,colortbl}
\usepackage{comment}
\usepackage[colorlinks=true, allcolors=blue]{hyperref}

\usepackage{tikz}
\usetikzlibrary{arrows,matrix,positioning}
\usepackage{pst-node}
\usepackage{easybmat}
\usepackage{dsfont}
\newcommand{\C}{{\mathbb C}}
\newcommand{\R}{{\mathbb R}}

\newcommand{\Hq}{{\mathbb H}}
\newcommand{\Oc}{{\mathbb O}}

\newcommand{\id}{{\mathbb I}}
\newcommand{\im}{{\rm i\,}}

\newcommand{\e}{{\bf e}}

\newcommand{\be}{\begin{eqnarray}}
\newcommand{\ee}{\end{eqnarray}}

\title{Notes on Spinors and Polyforms II: Quaternions and Octonions}
\author{Niren Bhoja and Kirill Krasnov\\ {}\\
{\it School of Mathematical Sciences, University of Nottingham, NG7 2RD, UK}}

\theoremstyle{definition}

\begin{document}
\maketitle

\begin{abstract} Pauli matrices are $2\times 2$ tracefree matrices with a real diagonal and complex (complex-conjugate) off-diagonal entries. They generate the Clifford algebra ${\rm Cl}(3)$. They can be generalised by replacing the off-diagonal complex number by one taking values in either quaternions or octonions (or their split versions).  These quaternionic and octonionic generalisations generate well-known models of ${\rm Cl}(5)$ and ${\rm Cl}(9)$ respectively. The main aim of the paper is to explicitly relate these models to the models arising via the creation/annihilation operator construction. We describe in details the models related to quaternions and octonions, as well as to the split quaternions and octonions. In particular, we record the description of the possible types of Weyl spinors of ${\rm Spin}(4,4)$, which does not seem to have appeared in the literature. 
\end{abstract}

\section{Introduction}

In an accompanying paper \cite{BK} we presented the construction of a set of creation/annihilation operator models for the Clifford algebra ${\rm Cl}(r,s), r+s=2m$. For some special values of $(r,s)$ the Clifford algebras admit a description in terms of quaternions and octonions (or their split versions). When such a description is available, it gives a formalism that is preferred in view of its compactness and computational power. However, some geometric aspects of spinors (for example the geometry related to pure spinors) are more manifest in the creation/annihilation operator formalism. For this reason, it is desirable to provide an explicit dictionary between the two formalisms, when they both exist. This will translate the geometric aspects that are manifest in the creation/annihilation operator picture to the quaternionic/octonionic picture. This is the main aim of the present paper. Thus, we provide an explicit dictionary between the description of ${\rm Cl}(4), {\rm Cl}(2,2), {\rm Cl}(8)$ and ${\rm Cl}(4,4)$ in terms of creation/annihilation operators and the description of the same Clifford algebras in terms of quaternions, split quaterions, octonions and split octonions respectively. The other aim is to record the possible types of orbits in the space of Weyl spinors and discuss the corresponding geometry. 

The creation/annihilation operator models are explained in the accompanying paper, and will be reviewed below (for the cases of relevance). The starting point of the quaternion/octonion description of Clifford algebras is the Pauli matrices description of ${\rm Cl}(3)$. Thus, let 
\be
\sigma^1 = \left(\begin{array}{cc} 0 & 1 \\ 1 & 0 \end{array}\right), \qquad \sigma^2 = \left(\begin{array}{cc} 0 & -\im \\ \im & 0 \end{array}\right), \qquad \sigma^3 = \left(\begin{array}{cc} 1 & 0 \\ 0 & -1 \end{array}\right)
\ee
be the usual Pauli matrices. These matrices anti-commute and square to plus the identity. Therefore, they generate the Clifford algebra ${\rm Cl}(3)\equiv{\rm Cl}(3,0)$, where our convention on ${\rm Cl}(r,s)$ is that this is the algebra generated by $\Gamma_I \Gamma_J + \Gamma_J \Gamma_I = 2 g_{IJ}$, where $g_{IJ}$ is a metric of signature $r$ plusses and $s$ minuses. 

We then note that a general linear combination of the Pauli matrices can be written as
\be\label{X}
X(r,q)=\left( \begin{array}{cc} r & \bar{q} \\ q & - r \end{array}\right),
\ee
with $q\in \C, r\in \R$ and $X(r,q)= {\rm Re}(q) \sigma^1 + {\rm Im}(q)\sigma^2 + r\sigma^3$. We now note that $q\in \C$ here can be replaced with $q\in \Hq$ or $q\in \Oc$, where $\Hq$ and $\Oc$ are the spaces of quaternions and octonions respectively. The corresponding matrices then generate ${\rm Cl}(5)$ and ${\rm Cl}(9)$ respectively. A subtlety arises in the case of octonions for these are non-associative. To deal with this, one replaces the operation of multiplication with $q$ with the operator $L_q$ of e.g. left multiplication. This is equivalent to a convention for the order of multiplication and deals with the issue of non-associativity. The arising description makes it clear that the spinors of ${\rm Cl}(3), {\rm Cl}(5), {\rm Cl}(9)$ are two-component columns with values in $\C,\Hq$ and $\Oc$ respectively. The described model of the Clifford algebras ${\rm Cl}(3), {\rm Cl}(5), {\rm Cl}(9)$ is not new and can be found in e.g. \cite{Harvey}. 

This construction can be further extended in several different ways. First, one can consider $r=0$ off-diagonal $\Gamma$-matrices. These generate the Clifford algebras ${\rm Cl}(2), {\rm Cl}(4), {\rm Cl}(8)$ for $q\in \C,\Hq,\Oc$ respectively. One can also choose a preferred vector in $\R^2, \R^4, \R^8$ and thus obtain a $\C, \Hq, \Oc$-based model for ${\rm Cl}(1), {\rm Cl}(3), {\rm Cl}(7)$ respectively. One can also apply the tensor product construction and generate $\Gamma$-matrices for ${\rm Cl}(3,1), {\rm Cl}(5,1), {\rm Cl}(9,1)$. These become $4\times 4$ matrices with entries in $\C,\Hq,\Oc$ respectively. We will describe the corresponding constructions in the main text. Finally, one can replace $\C,\Hq,\Oc$ with their split versions. The $X$-matrices (\ref{X}) then generate ${\rm Cl}(2,1), {\rm Cl}(3,2)$ and ${\rm Cl}(5,4)$ respectively and one can easily go either one dimension down or up staying in the same formalism. Overall, we get a useful for computations and economic description of a host of low (and not so low) dimensional Clifford algebras. 

It is desirable to relate the creation/annihilation operator construction and the described quaternion/octonion one, when they are both applicable. This is the aim of the present paper. We will see that each of the two constructions emphasises a different geometrical aspect of spinors in the corresponding dimension. And to fully understand what spinors are one needs to know both constructions, when they are both available. 

We also record the description of the various possible orbits of ${\rm Spin}(4,4)$ on the space of its Weyl spinors, which appears to be new. The space of ${\rm Spin}(4,4)$ Weyl spinors satisfying the reality condition (i.e. Majorana-Weyl spinors) can be identified with split octonions. Thus, Weyl spinors can be described as complexified split octonions. Null octonions (or null complexified octonions) correspond to pure spinors, and their classification is available from the general results on the classification of pure spinors, see in particular \cite{KT}. Majorana-Weyl spinors are also easy to classify. But we are not aware of any discussion of the possible types of orbits in the general case, i.e. complexified split octonions that are not null. In the analogous case of ${\rm Spin}(8)$ it is known that a general Weyl spinor can be used to construct a certain pure spinor. It thus follows that a general Weyl spinor continues to define a complex structure on $\R^8$ and its stabiliser is the same as that of a pure spinor, which is ${\rm SU}(4)$. We observe that this phenomenon continues to the ${\rm Spin}(4,4)$ setting, and that a general Weyl spinor that is of the form $\alpha+\im \beta, \alpha,\beta\in \Oc'$ with neither $\alpha$ nor $\beta$ null continues to define a certain pair of pure spinors. The possible stabilisers are then ${\rm SU}(2,2)$ when the corresponding pure spinor is of the type that defines a complex structure, and ${\rm SL}(4,\R)$ when the pair of the arising pure spinors defines a paracomplex structure. The new situation that has no ${\rm Spin}(8)$ analog is when either $\alpha$ or $\beta$ is null. This describes an orbit with interesting corresponding geometry, which we describe. 

Tee organisation of the paper is as follows. We proceed by describing in turn models related to quaternions in Section \ref{sec:spin-4}, split quaternions in \ref{sec:spin-22} and then octonions in and split octonions in Sections \ref{sec:spin-8} and \ref{sec:spin-44}. We conclude with a discussion. 

\section{Quaternionic models}
\label{sec:spin-4}

The aim of this Section is to see how the quaternionic description of ${\rm Spin}(4)$ arises within the creation/annihilation operator model. The Weyl spinors of ${\rm Spin}(4)$ are seen to be quaternions. We then describe how with similar methods one can describe ${\rm Cl}(6), {\rm Cl}(5,1), {\rm Cl}(4,2)$. Note that the corresponding Spin groups all contain ${\rm Spin}(4)$, and this is why the quaternionic description becomes possible. Only in the case ${\rm Cl}(5,1)$ the arising quaternionic description is sufficiently powerful to be useful. In the other two cases, while the quaternionic language is possible, they are most usefully described using the complex language. 

\subsection{Creation/annihilation operators construction of ${\rm Cl}(4)$}

We choose a complex structure on $\R^4$ thus identifying $\R^4=\C^2$. We will call the arising null complex coordinates $z_{1,2}$, and the corresponding one-forms $dz_{1,2}$. We introduce two pairs of creation/annihilation operators $a_{1,2}, a_{1,2}^\dagger$. The $\Gamma$ operators take the following form
\begin{equation}\label{gamma-spin4}
    \begin{split}
        \Gamma_4&=a_1+ a_1^{\dagger},\\
        \Gamma_2&=a_2 + a_2^{\dagger} ,\\
    \end{split}
    \qquad
    \begin{split}
        \Gamma_3&=-i(a_1 - a_1^{\dagger} ),\\
        \Gamma_1&=-i(a_2 - a_2^{\dagger} ).\\
    \end{split}
\end{equation}
We remind that in our conventions $a$ is the creation and $a^\dagger$ annihilation operator. The $a_{1,2}$ act as operators wedging the polyform they act on with $dz_{1,2}$ (from the left). The annihilation operators $a^\dagger_{1,2}$ look for a copy of $dz_{1,2}$ in the polyform and erase it. Some signs arise in this process, which can always be taken care of by rewriting the differential form one acts on so that the one-forms $dz_{1,2}$ one wants to kill are in the leftmost position. For more details, see the accompanying paper \cite{BK}. Note, however, that the notations in this paper are somewhat changed as compared to \cite{BK}. We have adopted the numbering and the signs in the imaginary $\Gamma$-matrices that become convenient below. A generic Dirac spinor (general polyform) is given by 
\begin{equation}\label{psi-spin4}
    \Psi=(u_1+u_2 dz_{12})+(v_1 dz_1+v_2 dz_2),
\end{equation}
where $dz_{12}:=dz_1\wedge dz_2$ and $u_{1,2},v_{1,2}\in\C$. In matrix notations, the Dirac spinor is a 4-component column. It is convenient to adopt the $2\times 2$ block notations, in which Weyl spinors are 2-component. Thus, we write
\be
\Psi = \left(\begin{array}{c} \psi_+ \\ \psi_- \end{array}\right), \qquad
\psi_+ = \left(\begin{array}{c} u_1 \\ u_2 \end{array}\right), \qquad
\psi_- = \left(\begin{array}{c} v_1 \\ v_2 \end{array}\right).
\ee
With these conventions, the $\Gamma$ operators can be written in the matrix notations
\be\label{gamma-matr-spin4}
\Gamma_4 =\left( \begin{array}{cc} 0 & \id \\ \id & 0 \end{array}\right), \quad 
\Gamma_i = \left( \begin{array}{cc} 0 & \im \sigma^i \\ -\im \sigma^i & 0 \end{array}\right), \quad i=1,2,3.
\ee
Here $\sigma^i$ are the usual Pauli matrices. It is this simple form of the resulting $\Gamma$-matrices that motivated the choices made in (\ref{gamma-spin4}), (\ref{psi-spin4}). 

The invariant inner product is determined by the following computation
\be
\langle \tilde{\Psi}, \Psi \rangle = ( \tilde{u}_1 - \tilde{u}_2 dz_{12} + \tilde{v}_1 dz_1 + \tilde{v}_2 dz_2) \wedge ( u_1 + u_2 dz_{12} + v_1 dz_1 + v_2 dz_2) \Big|_{top} = \\ \nonumber
(\tilde{u}_1 u_2-\tilde{u}_2 u_1) + (\tilde{v}_1 v_2 - \tilde{v}_2 v_1). 
\ee
The sign changes in the polyform $\tilde{\Psi}$ are understood as coming from rewriting this polyform with all decomposable summands written in the opposite order, see the accompanying paper. The inner product is thus an anti-symmetric pairing $\langle S_+, S_+\rangle, \langle S_-,S_-\rangle$. It can be written in matrix terms as
\be\label{inner-4}
\langle \Psi_1, \Psi_2\rangle = \Psi_1^T \left( \begin{array}{cc} \epsilon & 0 \\ 0 & \epsilon \end{array}\right) \Psi_2, \qquad \epsilon:= \im \sigma^2 =  \left( \begin{array}{cc} 0 & 1 \\ -1 & 0 \end{array}\right).
\ee
For the possible reality conditions, both $R=\Gamma_2 \Gamma_4 \ast$ and $R'=\Gamma_1 \Gamma_3\ast$ square to minus the identity, and so there are no Majorana spinors in this case. These two operators only differ by a sign in their action on $S_+$. Both of them can be used to define the hat operator. We have
\be\label{RRp-spin4}
R= \left( \begin{array}{cc} \epsilon & 0 \\ 0 & -\epsilon \end{array}\right)\ast, \qquad R'= \left( \begin{array}{cc} \epsilon & 0 \\ 0 & \epsilon \end{array}\right)\ast.
\ee
Thus, they differ by a sign of their action on $S_-$. We then define the hat operator
\be\label{hat}
\hat{\psi}_+ = \epsilon \left(\begin{array}{c} u_1 \\ u_2 \end{array}\right)^\ast =\left(\begin{array}{c} u_2^* \\ -u_1^* \end{array}\right) ,
\ee
which squares to minus the identity. Using the hat operator we have the following invariant norm on $S_+$
\be\label{norm-sp-spin4}
\langle R(\psi_+), \psi_+\rangle =  |u_1|^2+|u_2|^2.
\ee

\subsection{Quaternions and their complex description}

Quaternions $\Hq$ is a normed division algebra. The three imaginary unit quaternions satisfy 
\be
{\bf i}^2={\bf j}^2={\bf k}^2=-\id, \qquad {\bf i}{\bf j}={\bf k}.
\ee
A general quaternion is the linear combination
\be\label{quat}
q= q^4 \id + q^1 {\bf i} + q^2 {\bf j} + q^3 {\bf k}, \qquad q^{1,\ldots,4}\in \R.
\ee
The squared norm is given by
\be
|q|^2 = q\bar{q} = (q^4)^2 + \sum_{i=1}^3 (q^i)^2.
\ee 

Quaternions can be described in complex terms. To this end, one chooses a complex structure on $\Hq$, which identifies it with a copy of $\C^2$. There are $S^2$ worth of complex structures on $\R^4$, and a general complex structure can be described as the multiplication (left or right) by a unit imaginary quaternion. Let us choose the right multiplication $R_{\bf k}$ by ${\bf k}$ as the complex structure. The eigenspace of $R_{\bf k}$ of eigenvalue $-\im$ is spanned by $ \id + \im {\bf k}, {\bf i} - \im  {\bf j}$. This means that we can write a quaternion (\ref{quat}) as
\be\label{quat-complex-coord}
q(u_1,u_2) = u_1 (\id+\im {\bf k}) + u_2({\bf i} -\im {\bf j}) + u_1^* (\id-\im {\bf k}) + u_2^*({\bf i} +\im {\bf j}),
\ee
where
\be\label{uq}
u_1 = \frac{q_4-\im q_3}{2} , \qquad u_2 = \frac{q_1+\im q_2}{2}.
\ee
Note that the last two terms in (\ref{quat-complex-coord}) are the complex conjugates of the first two terms, where the complex conjugation reverses the sign in front of $\im$. The imaginary unit $\im$ should not be confused with the unit imaginary quaternion $\bf i$. Thus, with the complex structure $R_{\bf k}$ at hand, we get the isomorphism $\C^2\sim \Hq$ given by $q(u_1,u_2)$.

For later purposes we take two quaternions with complex coordinates $u_{1,2}$ and $\tilde{u}_{1,2}$ and compute the pairing
\be\label{uut-qqt}
\tilde{u}_1 u_2 - \tilde{u}_2 u_1 = \frac{1}{4} ( \tilde{q}_4 q_1 - \tilde{q}_1 q_4 + \tilde{q}_3 q_2 - \tilde{q}_2 q_3) + \frac{\im}{4} ( \tilde{q}_4 q_2 - \tilde{q}_2 q_4 + \tilde{q}_1 q_3 - \tilde{q}_3 q_1) = \frac{1}{4}(\tilde{q}, q (-{\bf i})) + \frac{\im}{4} (\tilde{q}, q (-{\bf j})).
\ee 
It is also interesting to see what various operations on $\C^2$ corresponds to in $\Hq$. We have
\be
q(u_2,-u_1)= q_1 \id- q_4 {\bf i} + q_3 {\bf j} - q_2 {\bf k} = (-{\bf i}) q(u_1, u_2), \\ \nonumber
q(u_1^*, u_2^*) = q_4 \id + q_1 {\bf i} - q_2 {\bf j} - q_3 {\bf k} = {\bf i} \,q(u_1, u_2)(-{\bf i}), 
\ee
and thus
\be\label{hat-quat}
q(u_2^*,-u_1^*) = q_1 \id - q_4 {\bf i} - q_3 {\bf j} +q_2 {\bf k} = q(u_1, u_2)(-{\bf i}) .
\ee
Thus, the anti-linear hat operation on $\C^2$ is just the right multiplication by $-{\bf i}$ when $\C^2$ is viewed as $\Hq$. We then have two operations on $\C^2$: the complex structure given by the multiplication by $-\im$, and the hat operator (\ref{hat}). On $\Hq$ the first is the right multiplication $R_{\bf k}$ and the second is the right multiplication $-R_{\bf i}$.  In their $\Hq$ versions these two operators clearly anti-commute. Having $R_{\bf i}$ and $R_{\bf k}$ we have the quaternionic structure on $\Hq$. This discussion shows how this is encoded in the $\C^2$ picture in the hat operator together with the operator of multiplication by $-\im$. 

We can now understand what the quaternion norm squared corresponds to in the complex description. Taking $\tilde{q}= q(u_1, u_2)(-{\bf i})$ in (\ref{uut-qqt}) we see that the imaginary term on the right-hand side $(q ({-\bf i}), q(-{\bf j}))= |q|^2({\bf i},{\bf j})=0$, and so drops out. What remains is 
\be
(|u_1|^2+|u_2|^2) =\frac{1}{4} |q|^2.
\ee

It is also interesting to see what the octonionic conjugation does as a map on $\C^2$. We see from (\ref{uq}) that
\be
(u_1,u_2)(\bar{q}) = ( u_1^*, - u_2)(q). 
\ee
So, the quaternion conjguation does not commute with the complex structure given by $R_{\bf k}$.

The left multiplication by an octonion $q$ commutes with $R_{\bf k}$. This means that it can be described as a $2\times 2$ complex matrix acting on the $-\im$ eigenspace of $R_{\bf k}$. A simple computation shows
\be
q( u_1 (\id+\im {\bf k}) + u_2({\bf i} -\im {\bf j})) = ((q^4-\im q^3)u_1 - (q^1-\im q^2)u_2) (\id+\im {\bf k}) + ((q^1+\im q^2)u_1 + (q^4+\im q^3)u_2) ({\bf i} - \im  {\bf j}).
\ee
This means we can describe the operator $L_q$ of left multiplication by $q$ as a $2\times 2$ matrix acting on the column $(u_1,u_2)$
\be
L_q \left( \begin{array}{c} u_1 \\ u_2 \end{array}\right) =   \left( \begin{array}{cc} q^4-\im q^3 & -(q^1-\im q^2) \\ q^1+\im q^2 & q^4+\im q^3 \end{array}\right)  \left( \begin{array}{c} u_1 \\ u_2 \end{array}\right) .
\ee
Thus, if we use the $\C^2$ description, quaternions can be identified with matrices of the type
\be\label{quat-matrix}
\left( \begin{array}{cc} a & -b^* \\ b & a^* \end{array}\right) , \qquad a,b\in \C,
\ee
with the left multiplication of quaternions described as multiplication of such matrices from the left. Quaternions can also be identified with 2-component complex-valued columns. This is e.g. the first column of the matrix (\ref{quat-matrix}). Note that the second column in (\ref{quat-matrix}) is the hat conjugate of the first, we see that the left multiplication by a quaternion has the same action on a 2-component column as on its hat conjugate. In other words, the left multiplication by a quaternion commutes with the hat conjugation. This is of course not surprising, because we already know that the hat conjugation corresponds to the right multiplication by $-{\bf i}$, which clearly commutes with $L_q$. 

\subsection{Spin(4) and quaternions}

We now go back to the $4\times 4$ $\Gamma$-matrices derived from the creation/annihilation operator construction. If we define 
\be
E^i := -\im \sigma^i
\ee
we have
\be
E^i E^j =- \delta^{ij} \id+  \epsilon^{ijk} E^k,
\ee
and so the objects $E^i$ can be identified with the unit quaternions ${\bf i,j,k}$. This means that we can rewrite the $\Gamma$-matrices (\ref{gamma-matr-spin4}) as 
\begin{equation}
    \Gamma_4=
        \begin{pmatrix}
            0&\id \\
            \id &0
        \end{pmatrix}\ ,\
    \Gamma_1=
        \begin{pmatrix}
            0&-{\bf i} \\
            {\bf i} &0
        \end{pmatrix}\ ,\
    \Gamma_2=
        \begin{pmatrix}
            0&-{\bf j}\\
            {\bf j}&0
        \end{pmatrix}\ ,\
    \Gamma_3=
        \begin{pmatrix}
            0&-{\bf k}\\
            {\bf k}&0
        \end{pmatrix}
\end{equation}
To put it differently, the general element of the Clifford algebra ${\rm Cl}(4)$ is
\be
X(q):= q^4 \Gamma_4 + q^i \Gamma_i = \begin{pmatrix}
            0&\bar{q}\\
            q &0
        \end{pmatrix},
        \ee
 where $q$ is a quaternion $\Hq\ni q= a^4\id + q^i E^i$. Thus, the matrix $X(q)$ is of the type (\ref{X}) with $r=0$. Such matrices act on two-component columns with entries in $\Hq$, and so semi-spinors of ${\rm Spin}(4)$ are quaternions $S_\pm =\Hq$. This shows how the quaternionic model of ${\rm Cl}(4)$ explained in the Introduction arises from the creation/annihilation operator model. 
 
 Let us compute the translation of the anti-linear map $R$ into the quaternion notation. From (\ref{RRp-spin4}) we see that it is given by the hat operator in the $\C^2$ description. From (\ref{hat-quat}) we see that it translates into the right multiplication by $-{\bf i}$ description. Thus, we can write
 \be
 R' = \left( \begin{array}{cc} -R_{\bf i} & 0 \\ 0 & -R_{\bf i} \end{array}\right),
 \ee
 which acts on $\Hq^2$. 
 
 It is also important to describe how the inner product (\ref{inner-4}) gets encoded in the quaternionic description. The inner product of two positive polyforms 
 \be
 \psi_+=u_1 + u_2 dz_{12}, \qquad \tilde{\psi}_+ = \tilde{u}_1 +  \tilde{u}_2 dz_{12}. 
 \ee
  is given by 
 \be
 \langle \tilde{\psi}_+, \psi_+\rangle = \left(\begin{array}{cc} \tilde{u}_1 & \tilde{u}_2 \end{array}\right) \left( \begin{array}{cc} 0 & 1 \\ -1 & 0 \end{array}\right)  \left(\begin{array}{c} u_1 \\ u_2 \end{array}\right) .
 \ee
Using (\ref{uut-qqt}) this translates into a complex quantity.
\be
4 \langle \tilde{\psi}_+, \psi_+\rangle =(\tilde{q}, q (-{\bf i})) + \im (\tilde{q}, q (-{\bf j})).
\ee
When $\tilde{\psi}_+=R'\psi_+$ we get 
\be
4 \langle R'\psi_+, \psi_+\rangle = |q|^2.
\ee
Thus, the inner product of a Weyl spinor with its hat conjugate translates into the norm squared of the corresponding quaternion. 
 
 \subsection{Quaternionic description of ${\rm Cl}(6)$}
 \label{sec:spin6}
 
 The fact that even/odd polyforms in $\C^2$ can be identified with quaternions allows us to give a quaternionic model of any Clifford algebra that acts on polyforms with a $\C^2$ factor. Let us see how this is possible for ${\rm Cl}(6)$. We start with the polyform description. 
 
 We choose a complex structure on $\R^6$, and introduce 3 complex null coordinates $z_{1,2,3}$, as well as the corresponding one-forms $dz_{1,2,3}$. We introduce 3 pairs of creation/annihilation operators $a_{1,2,3}, a_{1,2,3}^\dagger$. The Dirac spinor is a polyform which we choose to write as $\Psi=\psi_++\psi_-$ where
\be\label{polyform-6}
\psi_+ = u_1 + u_2 dz_{12}+ (v_1 dz_1 + v_2 dz_2)\wedge dz_3, \quad
\psi_- = (\tilde{u}_1 + \tilde{u}_2 dz_{12})\wedge dz_3+ \tilde{v}_1 dz_1 + \tilde{v}_2 dz_2.
\ee
The choices made here are motivated by the desire to have the Weyl spinors (\ref{psi-spin4}) of ${\rm Spin}(4)$ recognisable here. Thus, we see that $\psi_-$ is composed of a positive spinor $\tilde{u}_1 + \tilde{u}_2 dz_{12}$ and a negative spinor $\tilde{v}_1 dz_1 + \tilde{v}_2 dz_2$, and similarly for $\psi_+$.  The numbering of components in $\psi_+$ is such that the invariant product is a pairing $\langle S_+, S_-\rangle$ given by
\be\label{inner-spin6}
\langle \psi_-, \psi_+\rangle = \tilde{u}_1 u_2 -\tilde{u}_2 u_1 + \tilde{v}_1 v_2 - \tilde{v}_2 v_1
= \left( \begin{array}{cc} \tilde{u}_1 & \tilde{u}_2\end{array}\right) \epsilon \left( \begin{array}{c} u_1 \\ u_2\end{array}\right) + \left( \begin{array}{cc} \tilde{v}_1 & \tilde{v}_2\end{array}\right) \epsilon \left( \begin{array}{c} v_1 \\ v_2\end{array}\right).
\ee
We recognise this as being composed of the ${\rm Spin}(4)$ invariant product of ${\rm Spin}(4)$ Weyl spinors. 
The $\Gamma$ operators are given by
\begin{equation}\label{Gamma-6}
    \begin{split}
        \Gamma_4 &= a_1+a_1^{\dagger}, \\
        \Gamma_3&=- i(a_1-a_1^{\dagger}) 
    \end{split}
    \quad,\qquad
    \begin{split}
        \Gamma_2 &= a_2+a_2^{\dagger}, \\ 
        \Gamma_1&= -i(a_2-a_2^{\dagger})
    \end{split}
    \quad,\qquad
    \begin{split}
        \Gamma_5&= a_3+a_3^{\dagger}, \\ 
        \Gamma_6&= -i(a_3-a_3^{\dagger}),
    \end{split}
\end{equation}
where the numbering is to match that in (\ref{gamma-spin4}).
The anti-linear operator giving the reality condition is the product of 3 imaginary $\Gamma$-operators followed by the complex conjugation $R'=\Gamma_1\Gamma_3\Gamma_6\ast, (R')^2=\id$. 

We now identify the pairs of complex numbers $u_{1,2}, v_{1,2}$ and $\tilde{u}_{1,2}, \tilde{v}_{1,2}$ with 
quaternions that we call $u,v,\tilde{u},\tilde{v}$. We can thus write
\be\label{psi-spin6}
\psi_+ = u + v \wedge dz_3, \qquad \psi_- = \tilde{u} \wedge dz_3 + \tilde{v},
\ee
where it is understood that $u,\tilde{u}$ represent even polyforms and $v,\tilde{v}$ represent odd in $\Lambda(\C^2)$.

We place these quaternions into a 4-component column
\be
\Psi = \left(\begin{array}{c} \psi_+ \\ \psi_-\end{array}\right), \qquad \psi_+ = \left(\begin{array}{c} u \\ v \end{array}\right), \qquad \psi_- = \left(\begin{array}{c} \tilde{u} \\ \tilde{v} \end{array}\right).
\ee
This allows to describe the $\Gamma$-matrices $\Gamma_{1,2}, \Gamma_{4,5}$ in quaternionic terms as $\Gamma_p, p\in \Hq$. As we know from our discussion of the ${\rm Spin}(4)$ case, the operator $\Gamma_p$ maps an even quaternion $u$ to the odd quaternion $L_p u$, and an odd $v$ to the even $L_{\bar{p}} v$. Thus, in matrix notations
\be
\Gamma_p = \left( \begin{array}{cccc} 0 & 0 & 0 & L_{\bar{p}} \\ 0 & 0 & L_p & 0 
\\ 0 & L_{\bar{p}} & 0 & 0 \\ L_p & 0 & 0 & 0 \end{array}\right), \qquad p\in \Hq.
\ee
The other two $\Gamma$-matrices are worked out very easily using the (\ref{psi-spin6}) representation. We have
\be
\Gamma_3\psi_+ = u \wedge dz_3 - v, \qquad \Gamma_3 \psi_- = -\tilde{v} \wedge dz_3 + \tilde{u} , \\
\nonumber
\Gamma_6\psi_+ = -\im u \wedge dz_3 - \im v, \qquad \Gamma_6 \psi_- = \im \tilde{v} \wedge dz_3 + \im \tilde{u}.
\ee
This corresponds to the following $\Gamma$-matrices
\be
\Gamma_5 = \left( \begin{array}{cccc} 0 & 0 & 1 & 0 \\ 0 & 0 & 0 & -1 
\\ 1& 0 & 0 & 0 \\ 0 & -1 & 0 & 0 \end{array}\right), \qquad 
\Gamma_6 = \left( \begin{array}{cccc} 0 & 0 & \im & 0 \\ 0 & 0 & 0 & \im 
\\ -\im & 0 & 0 & 0 \\ 0 & -\im & 0 & 0 \end{array}\right).
\ee
We can rewrite these $\Gamma$-matrices in the $2\times 2$ form with 
\be\label{gamma1-5}
\Gamma_I = \left( \begin{array}{cc} 0 & \gamma_I \\ \gamma_I & 0 \end{array}\right), \qquad I=1,\ldots,5
\ee
where $\gamma_I$ are the generators of ${\rm Cl}(5)$ whose general linear combination is the matrix of the form (\ref{X}). To translate the last $\Gamma$-matrix into quaternionic notation, we need to use the fact that the multiplication by $-\im$ on $\C^2$ corresponds to the complex structure $R_{\bf k}$ on $\Hq$. This means that
\be\label{gamma6-spin6}
\Gamma_6 = \left( \begin{array}{cc} 0 & -R_{\bf k} \id \\ R_{\bf k} \id  & 0 \end{array}\right).
\ee
We note that using the quaternionic formalism leads to an economy of description. Thus, without quaternions the $\Gamma$-matrices are $8\times 8$ acting on $\C^8$. Using $\Hq=\C^2$ we have compacted them to $4\times 4$ matrices with quaternionic entries acting on $S=\Hq^4$. 

It is important to compute the $R'$ operator in the quaternionic notation. We start by computing it in the polyform notation, and then translate. Using (\ref{RRp-spin4}) we have the result for $\Gamma_1\Gamma_3\ast$, and so 
\be
R'=\Gamma_1\Gamma_3\Gamma_6\ast = (\Gamma_1\Gamma_3\ast) (-\Gamma_6) = \left( \begin{array}{cccc} \epsilon \ast& 0 & 0 & 0  \\ 0 & \epsilon \ast& 0 & 0 \\ 0 & 0 & \epsilon\ast & 0 \\0 & 0 & 0 & \epsilon \ast\end{array}\right) \left( \begin{array}{cccc} 0 & 0 & -\im & 0 \\ 0 & 0 & 0 & -\im 
\\ \im & 0 & 0 & 0 \\ 0 & \im & 0 & 0 \end{array}\right).
\ee
We then know that $\epsilon\ast$ is the hat operator whose quaternion translation is given by (\ref{hat-quat}), which is the right multiplication by $-{\bf i}$. The multiplication by $-\im$ is the action of the complex structure on $\C^2$, and this translates as $R_{\bf k}$. Thus, the quaternionic translation is
\be
R'= \left( \begin{array}{cc} 0 & - R_{\bf i} R_{\bf k} \id \\ R_{\bf i} R_{\bf k} \id & 0 \end{array}\right) =
\left( \begin{array}{cc} 0 & - R_{\bf j}  \id \\ R_{\bf j}  \id & 0 \end{array}\right) .
\ee
where the off-diagonal blocks are operators on $\Hq^2$. 

We can now see how to translate the invariant pairing (\ref{inner-spin6}) into quaternionic terms. We use (\ref{uut-qqt}) and have
\be\label{h2-pairing}
4\langle \psi_-, \psi_+\rangle = ( \tilde{u}, u(-{\bf i})) + ( \tilde{v}, v(-{\bf i})) + \im ( \tilde{u}, u(-{\bf j})) + \im ( \tilde{v}, v(-{\bf j})).
\ee
It is in general complex-valued. We can also see how an invariant norm of a Weyl spinor arises. We take $\psi_+$. We have $R'\psi_+\in S_-$, and the corresponding negative Weyl spinor is $(\tilde{u},\tilde{v})= 
(R_{\bf j} u , R_{\bf j} v)$, and so
\be
4\im \langle R'(\psi_+),\psi_+\rangle = |u|^2+ |v|^2.
\ee

It is instructive to compute the matrices representing the Lie algebra. The general Lie algebra element is 
\be
 \omega^{6I} \Gamma_6 \Gamma_I + \omega^{IJ} \Gamma_I \Gamma_J.
\ee
The products arising here are block-diagonal. The matrix $\omega^{IJ} \Gamma_I \Gamma_J$ has the same $S_\pm$ blocks given by
\be
 \left( \begin{array}{cc}  L_{\bf x} & L_{\overline{q}} \\ -L_q & L_{\bf y}\end{array}\right), \qquad {\bf x,y}\in {\rm Im}\,\Hq, \quad q\in\Hq.
 \ee
 The $S_+$ restriction of the matrix $\omega^{6I} \Gamma_6 \Gamma_I$ is
 \be
 (-R_{\bf k}) \left( \begin{array}{cc}  r & L_{\overline{p}} \\  L_p & - r \end{array}\right), \qquad r\in \R, \quad p\in\Hq.
 \ee
 The $S_\pm$ restrictions of the Lie algebra element can then be written as
 \be
 A_{\mathfrak{spin}(6)} = \left( \begin{array}{cc}  A-r R_{\bf k}  & -R_{\bf k} L_{\bar{p}} + L_{\overline{q}} \\ -R_{\bf k} L_p -L_q & A'+ rR_{\bf k} \end{array}\right), \quad
  A'_{\mathfrak{spin}(6)} = \left( \begin{array}{cc}  A+r R_{\bf k}  & R_{\bf k} L_{\bar{p}} + L_{\overline{q}} \\ R_{\bf k} L_p -L_q & A'- rR_{\bf k} \end{array}\right).
 \ee
 Here $A,A'$ are the two chiral $\mathfrak{su}(2)$-valued parts of the Lie algebra $\mathfrak{spin}(4)$. The presence of $R_{\bf k}$ in these expressions makes this quaternionic description of $\mathfrak{spin}(6)$ somewhat cumbersome. The purpose of the discussion was that it is possible. We note that we could similarly describe $\mathfrak{spin}(4,2)$ in quaternionic terms, but this description will also contain the operator $R_{\bf k}$ in the Lie algebra matrices, and so unlikely to be useful. The quaternionic formalism becomes most useful for the case ${\rm Cl}(5,1)$ which we now turn to. 
 
\subsection{Quaternionic description of ${\rm Cl}(5,1)$}

The described above quaternionic model of ${\rm Cl}(6)$ is plagued by the presence of the operator $R_{\bf k}$ encoding the multiplication by $-\im$ in the quaternionic language. This operator disappears from the $\Gamma$-matrices of ${\rm Cl}(5,1)$, which gives a much more powerful quaternionic description. In fact, we have ${\rm Spin}(5,1)={\rm SL}(2,\Hq)$. 

We first review the polyform model, and then provide its quaternionic translation. The only difference as compared to the ${\rm Cl}(6)$ case is that the third null coordinate is now real $dz_3\to du$. So,
we write the general polyform similar to (\ref{polyform-6}), but replacing $dz_3\to du$.
\be\label{polyform-51}
\psi_+ = u_1 + u_2 dz_{12}+ (v_1 dz_1 + v_2 dz_2)\wedge du, \quad
\psi_- = (\tilde{u}_1 + \tilde{u}_2 dz_{12})\wedge du+ \tilde{v}_1 dz_1 + \tilde{v}_2 dz_2.
\ee
All $\Gamma$-operators are unchanged as compared to what they are in the ${\rm Cl}(6)$ case, apart from the last $\Gamma$-operator is now $\Gamma_6 = -(b-b^\dagger)$ instead. The Weyl spinors are still objects in $\C^4$, and the invariant inner product is still given by (\ref{inner-spin6}). 

We can again identify $\C^2=\Hq$, and describe Weyl spinors as 2-component columns with entries in $\Hq$. The quaternionic version of the inner product is (\ref{h2-pairing}). In matrix notations, the first five $\Gamma$-matrices are still given by (\ref{gamma1-5}). The last $\Gamma$-matrix is now given by
\be\label{gamma-last}
\Gamma_6 = \left( \begin{array}{cc} 0 & \id \\ -\id & 0 \end{array}\right).
\ee

It is instructive to compute the matrices forming the Lie algebra. Taking products of distinct $\Gamma$-matrices we get a block-diagonal matrix with $2\times 2$ blocks on the diagonal. The upper-left block is the representation of $\mathfrak{spin}(5,1)$ on $S^+$, and the lower-right block gives the representation on $S^-$. We get
\be
A_{{\mathfrak spin}(5,1)} = \omega^{6I} \gamma_I + \omega^{IJ} \gamma_I \gamma_J, \qquad
A'_{{\mathfrak spin}(5,1)} = -\omega^{6I} \gamma_I + \omega^{IJ} \gamma_I \gamma_J,
\ee
A simple computation shows that these are matrices of the form
\be\nonumber
A_{{\mathfrak spin}(5,1)} = \left( \begin{array}{cc} r\id + {\bf x} & \overline{p+q} \\ p-q & -r\id + {\bf y}\end{array}\right), \qquad 
A'_{{\mathfrak spin}(5,1)} = \left( \begin{array}{cc} -r\id + {\bf x} & \overline{-p+q} \\ -p-q & r\id + {\bf y}\end{array}\right), \qquad 
r\in \R, p,q\in \Hq, {\bf x},{\bf y}\in {\rm Im}\, \Hq.
\ee
In other words, each is an arbitrary $2\times 2$ matrix with quaternionic entries satisfying the condition that the real part of its trace is zero. This is one way to argue that ${\rm Spin}(5,1)={\rm SL}(2,\Hq)$. We note that ${\rm SL}(2,\Hq)$ acts transitively on the space $\Hq^2$ of Weyl spinors. 

The possible anti-linear operators are $R=\Gamma_2\Gamma_4\Gamma_5\Gamma_6 \ast$ and $R'=\Gamma_1\Gamma_3 \ast$. Both of them now square to minus the identity, so there are no Majorana spinors in this signature. The $R'$ acts on all copies of $\C^2$ as $\epsilon\ast$. The operator $\epsilon\ast$ is the hat operator, which translated to quaternions becomes $-R_{\bf i}$. Thus, we can write
\be
R' \Big|_{S_+} = \left( \begin{array}{cc} -R_{\bf i}  & 0 \\ 0 & -R_{\bf i} \end{array}\right).
\ee
Both $R,R'$ operators preserve the helicity, which means that there is no quadratic invariant that can be constructed for a Weyl spinor in this signature, which agrees with the fact that the action of ${\rm Spin}(5,1)$ on $S_\pm$ is transitive. 

A positive Weyl spinor $\psi_+\in \Hq^2$ is pure. The geometric information it describes is that of a real null vector in $\R^{5,1}$, as well as a pair of two complex null vectors. The real null direction can be recovered by computing $\langle R'(\psi_+), \Gamma\psi_+\rangle$. Acting with $\Gamma_p$ on $\psi_+=(u,v)$ we have the spinor $(L_{\bar{p}} v, L_p u)\in S_-$. Using (\ref{h2-pairing}) we get
\be
4\langle \Gamma_p \psi_+, R'(\psi_+)\rangle = -(L_{\bar{p}} v, u) - (L_p u, v) + \im (L_{\bar{p}} v, u R_{\bf k} ) +\im (L_p u, v R_{\bf k}) = -2 (v, L_p u ).
\ee
Similarly, $\Gamma_5 \psi_+ = (u,-v)\in S_-$ and
\be
4\langle \Gamma_5 \psi_+, R'(\psi_+)\rangle = -|u|^2 +|v|^2 + \im (u, u R_{\bf k} ) -\im (v, v R_{\bf k}) = -|u|^2 +|v|^2.
\ee
Finally, we have $\Gamma_6\psi_+ = (u,v)\in S_-$ and 
\be
4\langle \Gamma_6 \psi_+, R'(\psi_+)\rangle = -|u|^2 -|v|^2 + \im (u, u R_{\bf k} ) +\im (v, v R_{\bf k}) = -|u|^2 -|v|^2.
\ee
Thus, overall we get
\be
-4\langle \Gamma \psi_+, R'(\psi_+)\rangle = ( 2 (v, L_p u ), |u|^2-|v|^2, |u|^2+|v|^2)\in \R^{1,5}
\ee
The right-hand side is a null vector. Taking a unit spinor $|u|^2+|v|^2=1$, the $\R^5$ part of this vector lies on $S^4$, which exhibits the quaternionic Hopf fibration $\Hq^2\ni S^7 \to S^4=\Hq$. To summarise, a quaternionic model is possible for any Spin group that contains ${\rm Spin}(4)$. It is however most useful in the case ${\rm Spin}(5,1)$ where we have ${\rm Spin}(5,1)={\rm SL}(2,\Hq)$.

\section{Models related to split quaternions}
\label{sec:spin-22}

The Clifford algebra ${\rm Cl}(2,2)$ is related to split quaternions. One can also obtain the split quaternion models of all Clifford algebras that contain ${\rm Cl}(2,2)$. Of these, the most interesting model that arises is that of ${\rm Cl}(3,3)$, because in this case one can identify ${\rm Spin}(3,3)={\rm SL}(2,\Hq')$. 

\subsection{Creation/annihilation operator model of ${\rm Spin}(2,2)$}

There are two possible models in this case. The model that exhibits the maximum analogy to the case of ${\rm Spin}(4)$, and also identifies semi-spinors with split quaternions is the one that endows $\R^{2,2}$ with a complex structure. Let $z_{1,2}$ be the corresponding complex coordinates. A general Dirac spinor is still given by the polyform (\ref{psi-spin4}).
The $\Gamma$-operators are given by
\begin{equation}\label{gamma-spin22}
    \begin{split}
        \Gamma_4&=a_1+ a_1^{\dagger}\\
        \Gamma_2&=-\im (a_2+a_2^{\dagger})
    \end{split}
    \quad,\qquad
    \begin{split}
        \Gamma_3&=-\im( a_1- a_1^{\dagger}) \\
        \Gamma_1&=a_2- a_2^{\dagger}.
    \end{split}
\end{equation}
Here $\Gamma_{4,3}$ are the same as in (\ref{gamma-spin4}). The other choices are motivated by a computation below. The $\Gamma$-matrices are then easily recoverable from (\ref{gamma-matr-spin4}) and are given by
\be\label{gamma-matr-spin22}
\Gamma_4= \left( \begin{array}{cc} 0 & \id \\ \id & 0 \end{array}\right), \quad \Gamma_3= \left( \begin{array}{cc} 0 & \im\sigma^3 \\ -\im\sigma^3 & 0 \end{array}\right), \quad \Gamma_1= \left( \begin{array}{cc} 0 & -\sigma^1 \\ \sigma^1 & 0 \end{array}\right), \quad \Gamma_2= \left( \begin{array}{cc} 0 & \sigma^2 \\ -\sigma^2 & 0 \end{array}\right).
\ee
The invariant inner product is still given by (\ref{inner-4}).

The two non-trivial anti-linear operators that can be constructed are $R= \Gamma_4\Gamma_1 \ast$ and $R'=\Gamma_2\Gamma_3 \ast$. We use $R'$ as the reality condition operator. In matrix form we have
\be
R' = \left(\begin{array}{cc} \sigma^1 & 0 \\ 0 & \sigma^1 \end{array} \right) \ast.
\ee
The action of $R'$ preserves $S_+$ (and $S_-$), and allows us to define Majorana-Weyl spinors. The action of $R$ on $S_+$ is the same as that of $R'$. The following invariant norm can be constructed
\be\label{spinor-norm-22}
\langle R'(\psi_+), \psi_+\rangle = |u_2|^2-|u_1|^2.
\ee

\subsection{Split quaternions and their complex description}

Quaternions $\Hq'$ is a normed composition algebra. The three imaginary unit quaternions $\bf{\tilde{i}}, \bf{\tilde{j}}, \bf{\tilde{k}}$ satisfy 
\be
\bf{\tilde{i}}^2=\bf{\tilde{j}}^2=\id, \qquad \bf{\tilde{k}}^2=-\id, \qquad \bf{\tilde{i}}\bf{\tilde{j}}=\bf{\tilde{k}}.
\ee
A general quaternion is the linear combination
\be
\tilde{q}= \tilde{q}^4 \id + \tilde{q}^1 {\bf\tilde{i}} + \tilde{q}^2 {\bf\tilde{j}}+ \tilde{q}^3 {\bf\tilde{k}}, \qquad \tilde{q}^{1,\ldots,4}\in \R.
\ee
The squared norm is given by
\be\label{norm-22}
|\tilde{q}|^2 = \tilde{q}\overline{\tilde{q}} = (\tilde{q}^4)^2 + (\tilde{q}^3)^2- (\tilde{q}^2)^2-(\tilde{q}^1)^2.
\ee 

Similarly to $\Hq$, the split quaternions $\Hq'$ can be described in complex terms. We want to select a complex structure on $\Hq'$ that would commute with the left multiplication. The only available choice now is $J=R_{\bf\tilde{k}}$.  The $-\im$ eigenspace of $R_{\bf k}$ is now spanned by $ \id + \im {\bf\tilde{k}}, {\bf\tilde{i}} + \im  {\bf\tilde{j}}$. This means that we can write a general split quaternion as
\be
q = u_1 ( \id + \im {\bf\tilde{k}}) + u_2 ({\bf\tilde{i}} + \im  {\bf\tilde{j}}) + 
u_1^* ( \id - \im {\bf\tilde{k}}) + u_2^* ({\bf\tilde{i}} - \im  {\bf\tilde{j}}),
\ee
where
\be
u_1= \frac{q_4 - \im q_3}{2}, \qquad u_2= \frac{q_1 - \im q_2}{2}.
\ee
The norm squared is then $|q|^2= 4|u_1|^2-4|u_2|^2$. 

Let us describe how the action of the complex conjugation and $\sigma^1 \ast$ get translated to $\Hq'$. We have
\be
q( u_1^*, u_2^*) =q^4 \id + q^1 {\bf\tilde{i}} - q^2 {\bf\tilde{j}}- q^3 {\bf\tilde{k}} = {\bf\tilde{i}} \,q( u_1, u_2) {\bf\tilde{i}}, \\ \nonumber
q( u_2, u_1) =q^1 \id + q^4 {\bf\tilde{i}} + q^3 {\bf\tilde{j}}+ q^2 {\bf\tilde{k}} = {\bf\tilde{i}}\, q( u_1, u_2) .
\ee
This means that the combination of these operations, which is the operation $\sigma^1\ast$, is given by the right multiplication by ${\bf\tilde{i}}$
\be
R' = R_{\bf \tilde{i}}.
\ee
Let us also work out what the inner product (\ref{inner-4}) becomes in the split quaternion notation. We have
\be
\tilde{u}_1 u_2 - \tilde{u}_2 u_1 = \frac{1}{4} (\tilde{q}_4 q_1 - \tilde{q}_3 q_2 - \tilde{q}_1 q_4 + \tilde{q}_2 q_3) +\frac{1}{4\im} (\tilde{q}_4 q_2 + \tilde{q}_3 q_1 - \tilde{q}_1 q_3 - \tilde{q}_2 q_4) = \frac{1}{4} (\tilde{q}, q{\bf\tilde{i}}) + \frac{1}{4\im} (\tilde{q},q{\bf\tilde{j}}) .
\ee
This shows how the inner product of two $\C^2$ vectors gets translated into the split quaternion notation. 

Let us understand what the left multiplication by a split quaternion corresponds to in the $\C^2$ notation. We have
\be\label{split-quat-pairing}
\tilde{q}( \alpha (\id+\im {\bf\tilde{k}}) + \beta({\bf\tilde{i}} +\im {\bf\tilde{j}})) = ((\tilde{q}^4-\im \tilde{q}^3)\alpha + (\tilde{q}^1+\im \tilde{q}^2)\beta) (\id+\im {\bf\tilde{k}}) + ((\tilde{q}^1-\im \tilde{q}^2)\alpha + (\tilde{q}^4+\im \tilde{q}^3)\beta) ({\bf\tilde{i}} + \im  {\bf\tilde{j}}).
\ee
We can again describe the operator $L_{\tilde{q}}$ of left multiplication by $\tilde{q}$ as a $2\times 2$ matrix acting on the column $(\alpha,\beta)$
\be
L_{\tilde{q}} \left( \begin{array}{c} \alpha \\ \beta \end{array}\right) =   \left( \begin{array}{cc} \tilde{q}^4-\im \tilde{q}^3 & \tilde{q}^1+\im \tilde{q}^2 \\ \tilde{q}^1-\im \tilde{q}^2 & \tilde{q}^4+\im \tilde{q}^3 \end{array}\right)  \left( \begin{array}{c} \alpha \\ \beta \end{array}\right) = ( \tilde{q}^4 \id + \tilde{q}^i \tilde{E}^i )\left( \begin{array}{c} \alpha \\ \beta \end{array}\right),
\ee
where
\be\label{E-split-quat}
\tilde{E}^1= \sigma^1, \quad \tilde{E}^2= -\sigma^2, \quad \tilde{E}^3=-\im \sigma^3.
\ee
Thus, split quaternions can be identified with matrices of the type
\be\label{quat-matrix-split}
\left( \begin{array}{cc} a & b^* \\ b & a^* \end{array}\right) , \qquad a,b\in \C.
\ee
Split quaternions can also be identified with 2-component complex columns, e.g. the first column of the above matrix. The determinant of the matrix (\ref{quat-matrix-split}) is the norm (\ref{norm-22}). 

\subsection{${\rm Spin}(2,2)$ and split quaternions}

Allowing the off-diagonal entries in (\ref{X}) to be split quaternions we get an $\Hq'$ model for ${\rm Cl}(3,2)$, and when $r=0$ for ${\rm Cl}(2,2)$. To see how the creation/annihilation operator model maps into this, we just need to put the $\Gamma$-matrices (\ref{gamma-matr-spin22}) into the form (\ref{X}). We note that the matrices (\ref{E-split-quat}) anti-commute and satisfy $\tilde{E}^1\tilde{E}^2=\tilde{E}^3$, as well as $(\tilde{E}^1)^2=(\tilde{E}^2)^2=\id$ and $(\tilde{E}^3)^2=-\id$. Thus, they satisfy all the properties of $\bf{\tilde{i}}, \bf{\tilde{j}}, \bf{\tilde{k}}$ and can be identified with them. Then the object 
\be
q = q^4 \id + q^i \tilde{E}^i 
\ee
can be identified as a split quaternion. Note that as a matrix it is a matrix of the type (\ref{quat-matrix-split}). The general linear combination of the $\Gamma$-matrices (\ref{gamma-matr-spin22}) is then a matrix of the type (\ref{X}) with $r=0$. The $\Gamma$-matrices (\ref{gamma-matr-spin22}) act on 4-component columns with complex entries. But we have seen that the 2-component columns can be identified with split quaternions. Thus, after $\Gamma$-matrices are identified with matrices of the form (\ref{X}), they act on 2-component columns with entries in $\Hq'$. In particular, semi-spinors are $S_\pm =\Hq'$. 

Let us also translate the inner product (\ref{spinor-norm-22}) to the quaternion notation. Using the fact that $R'$ acts as $R_{\bf\tilde{i}}$, and using (\ref{split-quat-pairing}) we get
\be
\langle R'(\psi_+), \psi_+\rangle = \frac{1}{4} ( q {\bf\tilde{i}}, q{\bf\tilde{i}}) + \frac{1}{4\im} ( q {\bf\tilde{i}}, q{\bf\tilde{j}}) = - \frac{1}{4} |q|^2,
\ee
which matches (\ref{spinor-norm-22}).

\subsection{Split quaternion model of ${\rm Cl}(3,3)$}

We can construct the split quaternion model of any Clifford algebra that contains ${\rm Cl}(2,2)$. In six dimensions the two cases that are covered are ${\rm Cl}(4,2)$ and ${\rm Cl}(3,3)$. The last case is particularly interesting, as it leads to the isomorphism ${\rm Spin}(3,3)={\rm SL}(2,\Hq')$. We only consider this case. 

To split quaternion model is obtained by considering polyforms in $\Lambda(\C^2\oplus \R)$. We write the general polyform in exactly the same way (\ref{polyform-51}) as we did in the case of ${\rm Cl}(5,1)$. The first 4 $\Gamma$-operators are then the same as in the ${\rm Cl}(2,2)$ case and are given by (\ref{gamma-spin22}). The remaining $\Gamma$-operators are
\be
\Gamma_5= b+b^\dagger, \qquad \Gamma_6 = - (b- b^\dagger).
\ee
Writing spinors as 8-component columns with entries in $\C$ we get the following $\Gamma$-matrices
\be
\Gamma_I = \left( \begin{array}{cc} 0 & \gamma_I \\ \gamma_I & 0 \end{array}\right), \quad I=1,\ldots, 5, \qquad \Gamma_6 = \left( \begin{array}{cc} 0 & \id \\ -\id & 0 \end{array}\right),
\ee
where 
\be\label{gamma-matr-spin33}
\gamma_5= \left( \begin{array}{cc} \id & 0 \\ 0 & -\id \end{array}\right), \quad
\gamma_4= \left( \begin{array}{cc} 0 & \id \\ \id & 0 \end{array}\right), \quad 
\gamma_3= \left( \begin{array}{cc} 0 & \im\sigma^3 \\ -\im\sigma^3 & 0 \end{array}\right), \quad \gamma_1= \left( \begin{array}{cc} 0 & -\sigma^1 \\ \sigma^1 & 0 \end{array}\right), \quad 
\gamma_2= \left( \begin{array}{cc} 0 & \sigma^2 \\ -\sigma^2 & 0 \end{array}\right).
\ee
We can now identify $\C^2=\Hq'$. Under this identification the matrices $\gamma_I$ become the canonical matrices of the type (\ref{X}), so that
\be
r \gamma_5+ \sum_{a=1}^4 q^a \gamma_a = \left( \begin{array}{cc} r\id & L_{\bar{q}} \\ L_q & -r\id \end{array}\right), \qquad q\in \Hq'.
\ee
Overall, we get the model that is precisely analogous to the quaternionic model of ${\rm Cl}(5,1)$, with the only difference being that quaternions are replaced by the split quaternions. The computation of the Lie algebra is also analogous, and shows that $\mathfrak{spin}(3,3)=\mathfrak{sl}(2,\Hq')$. 

The reality condition operator is the product of all the imaginary $\Gamma$-operators followed by the complex conjugation, and continues to be given by what it is in the ${\rm Cl}(2,2)$ case, which is $\sigma^1\ast$. In the split quaternion description this becomes $R_{\bf\tilde{i}}$. 

It is also interesting to discuss the arising spinor geometry. The general Weyl spinor is pure, and defines two complex and one real null directions. The real null direction can be again recovered as $\langle R'(\psi_+), \psi_+\rangle$. Taking the spinor to be unit, we get the split version of the quaternionic Hopf fibration. 

We can also take the Weyl spinor to be null. In particular, Majorana-Weyl spinors are null. A Majorana-Weyl spinor describes three real null directions, which can be recovered by computing $\langle \psi_+, \Gamma\Gamma\Gamma\psi_+\rangle$. This 3-form in $\R^{3,3}$ is decomposable and is given by the product of the 3 null real directions when $\psi_+$ is Majorana-Weyl. 

\section{The case of ${\rm Spin}(8)$}
\label{sec:spin-8}

In the already described cases of ${\rm Cl}(4)$ and ${\rm Cl}(2,2)$ semi-spinors are 2-component columns with complex entries, and these can be identified with $\Hq,\Hq'$ respectively. The difference with the cases of ${\rm Cl}(8)$, ${\rm Cl}(4,4)$ is that it is the Majorana-Weyl spinors that can be identified with $\Oc,\Oc'$. Thus, the identification is only possible after a reality condition is imposed. To see how this works we first need to describe the Majorana-Weyl spinors explicitly.

\subsection{Creation/annihilation operator construction}

To construct ${\rm Cl}(8)$ we choose a complex structure, which identifies $\R^8\sim \C^4$. We denote the complex coordinates by $z_{1,\ldots,4}$, and the coordinate one-forms by $dz_{1,\ldots,4}$. We introduce four pairs of creation/annihilation operators $a_I, a_I^\dagger, I=1,\ldots, 4$. We introduce the following $\Gamma$-operators
\be\label{gamma-spin8}
\Gamma_{4+I} := a_I+ a_I^{\dagger} , \qquad \Gamma_{I} := -\im (a_I- a_I^{\dagger} ), \quad i=1,2,3,4.
\ee
Spinors are polyforms, i.e. elements of $\Lambda(\C^4)$, in general with complex coefficients. Weyl spinors are even or odd degree polyforms. Note that we have chosen the same signs as in (\ref{gamma-spin4}). 

The interesting anti-linear operator that arises in this case is given by the product of all four imaginary $\Gamma$-operators with the complex conjugation
\be
R := \Gamma^1 \Gamma^2 \Gamma^3 \Gamma^4 *,
\ee
It is easy to check that $R$ commutes with all the gamma-matrices. It is also easy to check that $R^2=\id$, so $R$ is a real structure. Since $R$ is composed of an even number of gamma-matrices, it preserves the spaces of Weyl spinors. Thus, it allows us to define the notion of Majorana-Weyl spinors for ${\rm Spin}(8)$. A simple computation shows that the action of $R$ is complex conjugation followed by the Hodge star in $\C^4$, modulo some signs. 
 
 \subsection{Majorana-Weyl spinors explicitly}
 
It is now convenient to choose a basis $e^I=dz_I$ of basic one-forms. Then a general odd/even polyform that is also real can be written as follows
 \be\label{real-polyforms}
 \psi^- \equiv \psi^-(u)= u_1 e^1 + \bar{u}_1 e^{234} + u_2 e^2 + \bar{u}_2 e^{314}+ u_3 e^3 +\bar{u}_3 e^{124} +\im \bar{u}_4 e^4 +  \im u_4 e^{123}, \\ \nonumber
 \psi^+ \equiv \psi^+(u) = 
  u_1 e^{41}+\bar{u}_1 e^{23} +u_2 e^{42}+\bar{u}_2 e^{31} +u_3 e^{43}+\bar{u}_3 e^{12} +\im \bar{u}_4+\im u_4 e^{4123}  ,
 \ee
 where the quantities $u_I$, with $I=1,\ldots,4$ are complex numbers. This particular choice of the complex coordinates $u_I$, and in particular the choices made for the coordinate $e^4$, will become justified below by the desired form of the action of the $\Gamma$-matrices.

The inner product is a pairing $\langle S_+, S_+\rangle, \langle S_-, S_-\rangle$. If we take two positive spinors $\psi^+(\tilde{u}), \psi^+(u)$, the product $\langle \psi^+(\tilde{u}), \psi^+(u)\rangle$ is computed by taking the polyform $\psi^+(\tilde{u})$ in the reverse order, wedging with $\psi^+(u)$ and projecting on the top component. A simple computation gives
\be
\langle \psi^+(\tilde{u}), \psi^+(u)\rangle= 2{\rm Re}\sum_{I=1}^4  \tilde{u}_{I} \bar{u}_{I}.
\ee
Thus, Majorana-Weyl spinors are identified $S_\pm \sim \C^4$, with the invariant pairings on $S_\pm$ being given by the standard definite Hermitian metric on $\C^4$. 

The form of the inner product makes it clear that the basis $e^{42}, e^{42}, e^{43}, \id$ of $S^+$ in (\ref{real-polyforms}) is totally null. To make contact with octonions that are usually described in a non-null basis, it is necessary to switch to a different parametrisation of polyforms. We parametrise the polyforms by the real and imaginary parts of $u_I$ and write
\be\label{u-ab}
u_1= \alpha_1+\im  \alpha_5, \quad u_2 =  \alpha_2 + \im  \alpha_6, \quad u_3=  \alpha_3+\im  \alpha_7, \quad u_4= \alpha_0 + \im  \alpha_4.
\ee
We will denote the components of the positive polyform by $\alpha$ and the negative polyform components by $\beta$. We then have
\be
\psi^+ = \alpha_1( e^{41}+e^{23}) + \alpha_2( e^{42}+e^{31}) +\alpha_3( e^{43}+e^{12}) + \alpha_4 (1- e^{4123}) \\
\nonumber
+ \im \alpha_5( e^{41}-e^{23}) + \im \alpha_6( e^{42}-e^{31}) +\im \alpha_7( e^{43}-e^{12}) + \im \alpha_0 (1+ e^{4123}) ,
\\ 
\psi^- = \beta_1( e^{1}+e^{423}) + \beta_2( e^{2}+e^{431}) +\beta_3( e^{3}+e^{412}) + \beta_4 (e^4- e^{123}) \\
\nonumber
+ \im \beta_5( e^{1}-e^{423}) + \im \beta_6( e^{2}-e^{431}) +\im \beta_7( e^{3}-e^{412}) + \im \beta_0 (e^4+ e^{123}).
\ee

 \subsection{The action of $\Gamma$-matrices}
 
 We now introduce a 16-component column 
 \be
\Psi = \left( \begin{array}{c} \psi^+ \\ \psi^-\end{array}\right) = \left( \begin{array}{c} \alpha_0 \\ \alpha_1 \\ \vdots \\ \alpha_7 \\ \beta_0 \\ \beta_1 \\ \vdots \\ \beta_7  \end{array}\right).
\ee
A computation shows that the $\Gamma$-operators become the following $16\times 16$ matrices
\begin{equation}\label{Gamma-matr-spin8}
    \Gamma_0=
    \begin{pmatrix}
    0&\id \\
    \id &0
    \end{pmatrix}
    , \quad
    \Gamma_a=
    \begin{pmatrix}
    0&-E_a\\
    E_a&0
    \end{pmatrix}
    \ \text{for} \ a\in \{1,\ldots,7\},
\end{equation}
where
\begin{equation}\label{E-matr-spin8}
    \begin{split} 
        E_1&=-E_{01}+E_{27}-E_{36}+E_{45},\\
        E_2&=-E_{02}-E_{17}+E_{35}+E_{46},\\
        E_3&=-E_{03}+E_{16}-E_{25}+E_{47},\\
        E_4&=-E_{04}-E_{15}-E_{26}-E_{37},\\
        E_5&=-E_{05}+E_{14}+E_{23}-E_{67},\\
        E_6&=-E_{06}-E_{13}+E_{24}+E_{57},\\
        E_7&=-E_{07}+E_{12}+E_{34}-E_{56}.
    \end{split}
\end{equation}

 \subsection{Octonions}
 
 The space of octonions $\Oc$ is a normed algebra with the property $|xy|=|x| |y|$ (i.e. a composition algebra). The usual octonions (unlike split octonions) also have the property that the norm of every non-zero element is not zero, which makes them into a division algebra. It is non-commutative and non-associative, but alternative, which can be stated as the property that the subalgebra generated by any two elements is associative. 
 
 A general octonion is an object
 \be
 q= q_0 \id +\sum_{a=1}^7 q_a {\bf e}^a,
 \ee
 where ${\bf e}^a$ are unit imaginary octonions. The unit octonions anti-commute and square to minus the identity. The octonion conjugate changes the sign of all the imaginary octonions. The octonionic pairing is
 \be
 (q,q)=|q|^2= q\overline{q} = (q_0)^2 + \sum_{a=1}^7 (q_a)^2.
 \ee
 
 We encode the octonionic product by the cross-product in the space of imaginary octonions. Thus, we write
 \be
 \Oc = \R + {\rm Im}\,\Oc.
 \ee
 Let $\e^{1,\ldots,7}$ be a basis in the space of imaginary octonions. The cross-product in ${\rm Im}\,\Oc$ can be encoded by the following 3-form
 \be\label{3-form}
C = \e^{567} + \e^5( \e^{41}-\e^{23}) + \e^6(\e^{42}-\e^{31}) + \e^7(\e^{43}-\e^{12}).
 \ee
 This encodes the cross-product in the sense that $C(\e^a, \e^b, \e^c)=( \e^a\times \e^b, \e^c)$, where the standard metric on $\R^7$ is used. So, for instance $\e^5 \times \e^6= \e^7$. 
 
  \subsection{Octonionic model for ${\rm Cl}(8)$}
 
 The octonionic product can also be encoded into $8\times 8$ matrices. To this end we represent a general octonion as an 8-component column. Then the operators of left multiplication by unit imaginary octonions can be checked to be given by $L_{{\bf e}^a}= E_a$, where $E_a$ are precisely the same matrices already encountered in (\ref{E-matr-spin8}). Coming back to our model for the Clifford algebra ${\rm Cl}(8)$ we see that the general linear combination of the $\Gamma$-matrices is
 \be\label{cl8-oct}
 q_0 \Gamma_0 + \sum_{a=1}^7 q_a \Gamma_a = \left( \begin{array}{cc} 0 & L_{\bar{q}} \\ L_q & 0 \end{array}\right). 
 \ee
 These are matrices of the type (\ref{X}) with $r=0$, which reproduces the model explained in the Introduction for $q\in \Oc$. We thus see that ${\rm Cl}(8)$ is generated by matrices (\ref{cl8-oct}) that act on 2-component columns with entries in $\Oc$. Majorana-Weyl spinors are then identified with copies of $\Oc$.

\subsection{Pure spinors}

In the creation/annihilation operator model, the pure spinors are decomposable polyforms. In particular, the model comes with two preferred pure spinors, the identity polyform, and the top polyform. For ${\rm Cl}(8)$, both are in $S_+$, and both are null spinors. It is also clear from (\ref{real-polyforms}) that only the linear combinations $\im (1+e^{4123})$ and $1-e^{4123}$ are Majorana-Weyl spinors. Majorana-Weyl spinors can be identified with $\Oc$, and are never null. Thus, we cannot see pure spinors if we restrict our attention to Majorana-Weyl spinors. 

In order to describe pure spinors we need arbitrary, not necessarily satisfying the reality condition polyforms. These can be described as complexified real polyforms, which are then identified with complexified octonions. Thus, in the octonionic description of ${\rm Cl}(8)$ the general Weyl spinor is a complexified octonion. In particular, pure spinors are necessarily complexified octonions, because they are null. 

Let us see how this works for the two canonical pure spinors. The Majorana-Weyl identity and ${\bf e}^e$ octonion spinors are
\be
\Oc\ni \id = \im (1+e^{4123}), \qquad \Oc\ni {\bf u}:={\bf e}^4= 1-e^{4123}.
\ee
Note that we have introduced a convenient notation and denoted ${\bf u}$ the imaginary unit octonion ${\bf e}^4$. This means that the pure spinors $1,e^{4123}\in S_+$ are given by
\be\label{spin8-can-pure}
1= \frac{1}{2\im} (\id+\im {\bf u}) , \qquad e^{4123}= \frac{1}{2\im} (\id-\im {\bf u}).
\ee
We have the operation of complex conjugation that reverses the sign in front of $\im$. This operation is not to be confused with the octonion conjugation. It is clear that the two polyforms $1,e^{4123}$ are related by the complex conjugation. This complex conjugation is the translation of the $R$ anti-linear operator that acts on polyform into the octonion picture. 

\subsection{The complex structure defined by a pure spinor}

By Cartan's general result pure spinors of ${\rm Spin}(2n)$ are Weyl spinors that have the property that all $B_k(\psi_+,\psi_+)$ are vanishing apart from $k=n$. For ${\rm Spin}(8)$ the only possible number of insertions of $\Gamma$-matrices between two Weyl spinors of the same helicity is zero, two and four. The product of two $\Gamma$-matrices restricted to $S_+$ is expressible as either $E_a$ or $E_a E_b$ with $a,b=1,\ldots, 7$. Both are anti-symmetric. Thus, $B_2(\psi_+,\psi_+)=0$. This shows that null spinors with $B_0(\psi_+,\psi_+)=0$ are pure. Because the inner product of a Majorana-Weyl spinor with itself is the norm squared of the corresponding octonion, the Majorana-Weyl spinors are not pure. 

In (\ref{spin8-can-pure}) we have an example of two canonical pure spinors. In the octonion description they become complex linear combinations of two unit octonions. From general considerations we know that either of these two pure spinors defines a complex structure in $\R^8$ and that $B_4(\psi_+,\psi_+)$ is the product of four null directions spanning one of the eigenspaces of this complex structure. It is interesting to compute both the complex structure and $B_4(\psi_+,\psi_+)$ explicitly. 

The complex structure defined by the pure spinor $\psi_+=1$ is easiest computed by computing $B_2(R(\psi_+), \psi_+)$. We have $R(1)\sim e^{4123}$ and so we need to compute
\be
B_2(R(\psi_+), \psi_+)\sim \langle(\id+\im {\bf u}), \Gamma\Gamma (\id- \im {\bf u})\rangle. 
\ee
The components of this 2-form in the $\id, x\in {\rm Im}(\Oc)$ directions are 
\be
\langle(\id+\im {\bf u}), \Gamma_\id \Gamma_x (\id- \im {\bf u})\rangle = ((\id+\im {\bf u}), L_x (\id-\im {\bf u})), \\ \nonumber
\langle(\id+\im {\bf u}), \Gamma_x \Gamma_y (\id- \im {\bf u})\rangle = ((\id+\im {\bf u}), L_{\bar{x}} L_y (\id-\im {\bf u})).
\ee
We also have 
\be
( (\id+\im {\bf u}), L_x (\id-\im {\bf u})) =2\im {\bf u}, \\ \nonumber
((\id+\im {\bf u}), L_x L_y (\id-\im {\bf u})) =2\im (e^{15} + e^{26} + e^{37}) : = 2\im \omega.
\ee
Thus, overall, we have the following 2-form in $\R^8$
\be\label{spin8-2form}
\frac{1}{2\im} ((\id+\im {\bf u}), \Gamma\Gamma (\id- \im {\bf u}))= \id\wedge {\bf u} - \omega.
\ee
This is the $(1,1)$ form for the arising complex structure, which allows us to recover the complex structure (e.g. by raising one of the indices on the two-form and thus interpreting it as an endomorphism of $\R^8$). 

To see what the complex structure is explicitly, it helps to start by considering the complex structure on $\Oc$ given by $L_{\bf u}$, i.e. the left multiplication by the unit imaginary octonion $\bf u$. The eigenvectors of eigenvalue $-\im$ are given by
\be
\id + \im {\bf u}, \qquad {\bf e}^1 + \im {\bf e}^5, \qquad {\bf e}^2 + \im {\bf e}^6, \qquad {\bf e}^3 + \im {\bf e}^7.
\ee
The K\"ahler form arising as $(\im/2) dz\wedge d\bar{z}$ is $\id\wedge {\bf u} + \omega$, which is (\ref{spin8-2form}) up to the sign in front of the last term. This tells us that we need to consider a slightly different complex structure on $\Oc$ to get (\ref{spin8-2form}). Let us instead consider the operator $R_{\bf u}$ of right multiplication by ${\bf u}$. Its $-\im$ eigenvectors are now
\be\label{directions-Ru}
\id + \im {\bf u}, \qquad {\bf e}^1 - \im {\bf e}^5, \qquad {\bf e}^2 - \im {\bf e}^6, \qquad {\bf e}^3 - \im {\bf e}^7,
\ee
and the corresponding K\"ahler form is precisely (\ref{spin8-2form}).

We can alternatively recover the complex structure with its $(0,1)$ and $(1,0)$ directions by computing $B_4(\psi_+,\psi_+)$ or $B_4(R(\psi_+),R(\psi_+))$. We know that both are decomposable and are given by the product of four $(0,1)$ null directions, or four $(1,0)$ directions. 

Thus, we want to compute
\be
\langle(\id-\im {\bf u}), \Gamma\Gamma\Gamma\Gamma (\id- \im {\bf u})\rangle \in \Lambda^4(\R^8).
\ee
The various components of this 4-form that we need are given by
\be\label{4-form-comps}
\langle\psi , \Gamma_\id \Gamma_x\Gamma_y\Gamma_z \psi\rangle=
(\psi, L_x L_{\bar{y}} L_z \psi), \\ \nonumber
\langle\psi, \Gamma_x \Gamma_y \Gamma_z\Gamma_w \psi\rangle=
(\psi, L_x L_{\bar{y}} L_z L_{\bar{w}}  \psi),
\ee
where on the right-hand side the spinor $\psi$ is interpreted as a (complexified) octonion. 
A computation gives
\be
((\id-\im{\bf u}), L_x L_y L_z (\id-\im {\bf u})) =  2\im \Omega, 
\\ \nonumber
((\id-\im{\bf u}), L_x L_y L_z L_w (\id-\im {\bf u})) = 2{\bf u}\wedge \Omega,
\ee
where
\be
\Omega := (e^1-\im e^5)\wedge (e^2-\im e^6) \wedge (e^3-\im e^7).
\ee
This means that
\be\label{1u-omega}
\frac{\im}{2} \langle(\id-\im {\bf u}), \Gamma\Gamma\Gamma\Gamma (\id- \im {\bf u})\rangle = ( \id+\im {\bf u}) \wedge \Omega.
\ee
This is precisely the wedge product of the directions (\ref{directions-Ru}), as expected. The 4-form obtained as $\langle(\id+\im {\bf u}), \Gamma\Gamma\Gamma\Gamma (\id+ \im {\bf u})\rangle$ is given by the product of the complex conjugate directions.

To summarise, we learn that the complex structure on $\R^8\sim \Oc$ that corresponds to the complex conjugate pair of pure spinors $\id+\im {\bf u}, \id-\im {\bf u}$ is given by $R_{\bf u}$, the right multiplication by ${\bf u}$. 

 \subsection{Majorana spinors}
 
Now that we understand the octonionic description of pure spinors, we can come back to Majorana-Weyl spinors. We already know that such spinors cannot be pure. In fact, we see from (\ref{spin8-can-pure}) that they are given by a linear combination of two pure spinors. 

 It is important to discuss the geometry of a Majorana-Weyl spinor $\psi_M\in S_+$. First, such a spinor has a non-vanishing norm $\langle \psi_M,\psi_M\rangle$, which coincides with the norm squared of the corresponding octonion. Second, the group ${\rm Spin}(8)$ acts on the space of Majorana-Weyl spinors of fixed norm and of one helicity transitively, with the stabiliser ${\rm Spin}(7)$. So, we have 
 \be
 S^7 ={\rm Spin}(8)/{\rm Spin}(7).
 \ee
This can be seen from the fact that the Majorana-Weyl representation is isomorphic to the vector representation $S_+\sim \R^8$. 

A Majorana-Weyl spinor of ${\rm Spin}(8)$ thus has ${\rm Spin}(7)$ as the stabiliser, and endows $\R^8$ with a ${\rm Spin}(7)$ structure. This is the 4-form $B_4(\psi_M,\psi_M)$, whose stabiliser in ${\rm GL}(8,\R)$ is ${\rm Spin}(7)$. It is instructive to compute this 4-form explicitly for $\psi_M=\id$. Any unit Majorana-Weyl spinor is in the ${\rm Spin}(8)$ orbit of this spinor. The components of this 4-form are given by (\ref{4-form-comps}), and we have
\be
(\id, L_x L_y L_z \id) = -C, \\ \nonumber
(\id, L_x L_y L_z L_w \id) = -{}^*C,
\ee
where 
\be
C:= e^{567} + e^5(e^{41}-e^{23}) + e^6( e^{42}-e^{31}) + e^7(e^{43}-e^{12}),
\\ \nonumber
{}^*C := e^{1234} + e^{67} (e^{41}-e^{23})+ e^{75} ( e^{42}-e^{31})+ e^{56} (e^{43}-e^{12}).
\ee
Note that $C$ is just the 3-form (\ref{3-form}) that encodes the octonion product, and ${}^*C$ is its Hodge dual in $\R^7$. Thus, overall, we have
\be\label{4-form}
\Lambda^4(\R^8) \ni \langle \id, \Gamma\Gamma\Gamma\Gamma\id\rangle = \id\wedge C - {}^* C.
\ee
We note, for future use, that
\be\label{C-Om}
C= {\bf u}\wedge \omega + {\rm Im}(\Omega), \qquad {}^*C = {\rm Re}(\Omega)\wedge {\bf u} + \frac{1}{2}\omega\omega.
\ee

\subsection{A general spinor}

A general Weyl spinor of ${\rm Spin}(8)$ is a complexified octonion. When the spinor is not null, we can always rescale it by a complex number to make it unit. As is explained in \cite{Charlton}, see page 33, a complex impure spinor defines a certain pure spinor. Indeed, assuming $\psi$ is unit and denoting $\lambda = \langle R(\psi), \psi\rangle$, consider
\be\label{tilde-psi-so8}
\tilde{\psi} = \frac{\lambda \psi - R(\psi)}{\sqrt{\lambda^2-1}}.
\ee
Then $\langle \tilde{\psi},\tilde{\psi} \rangle =-1$ and 
\be
\psi+\tilde{\psi}
\ee
is null and therefore pure. Because the stabiliser of $R(\psi)$ is the same as the stabiliser of $\psi$ (because the stabiliser is real), this stabiliser also coincides with that of $\tilde{\psi}$ and thus $\psi+\tilde{\psi}$. This means that the stabiliser of $\psi$ is that of a pure spinor, which is ${\rm SU}(4)$.

The above discussion suggests that a general complex spinor continues to define a complex structure on $\R^8$. Let us see how this arises. A complex spinor has a well-defined real and imaginary parts $\psi=\alpha + \im \beta, \alpha,\beta\in \Oc$. There are two invariant scalars that can be constructed
\be\label{psi-alpha-beta}
\langle \psi, \psi\rangle = |\alpha|^2-|\beta|^2+ 2\im (\alpha,\beta), \\ \nonumber
\langle R(\psi), \psi\rangle = |\alpha|^2+|\beta|^2.
\ee
Rescaling the spinor to make it unit, we assume $|\alpha|^2-|\beta|^2=1, (\alpha,\beta)=0$. Also, using the action of ${\rm Spin}(8)$ we can make $\alpha$ a multiple of $\id$, and then use ${\rm Spin}(7)$ that stabilises $\id$ to make $\beta$ (which is orthogonal to $\alpha$) to be a multiple of ${\bf u}$. This leads us to consider the unit spinor
\be
\psi = \cosh\tau \id + \im \sinh\tau {\bf u}.
\ee
Then $\lambda = \cosh^2\tau + \sinh^2\tau = \cosh(2\tau)$ and
\be
\tilde{\psi}  = \sinh\tau \id + \im \cosh\tau {\bf u},
\ee
so that
\be
\psi+\tilde{\psi} = (\cosh\tau+\sinh\tau)(\id + \im {\bf u}),
\ee
which is the pure spinor we already considered above. And indeed, we have the following results
\be
( (\cosh\tau \id+\im \sinh\tau {\bf u}), L_x (\cosh\tau \id-\im \sinh\tau {\bf u})) =\im \sinh(2\tau) {\bf u}, \\ \nonumber
((\cosh\tau \id+\im \sinh\tau {\bf u}), L_x L_y (\cosh\tau \id-\im \sinh\tau {\bf u})) =\im \sinh(2\tau) \omega,
\ee
which shows that
\be
\langle (\cosh\tau \id+\im \sinh\tau {\bf u}), \Gamma\Gamma (\cosh\tau \id-\im \sinh\tau {\bf u}) = \im\sinh(2\tau) ( \id\wedge {\bf u} - \omega).
\ee
This is a multiple of the K\"ahler 2-form for the complex structure defined by the pure spinor $\id+\im {\bf u}$. 

To compute the 4-form $B_4(\psi,\psi)$ we need the following results
\be\nonumber
((\cosh\tau \id-\im \sinh\tau {\bf u}), L_x L_y L_z  (\cosh\tau \id-\im \sinh\tau {\bf u})) =\im \sinh(2\tau) {\rm Re}(\Omega) - \cosh(2\tau) {\rm Im}(\Omega)- {\bf u}\wedge \omega, \\ \nonumber
((\cosh\tau \id-\im \sinh\tau {\bf u}), L_x L_y L_z  L_w(\cosh\tau \id-\im \sinh\tau {\bf u})) =- \cosh(2\tau) {\rm Re}(\Omega)\wedge{\bf u} - \im \sinh(2\tau) {\rm Im}(\Omega)\wedge{\bf u} - \frac{1}{2}\omega \wedge \omega.
\ee
This means that
\be
\langle (\cosh\tau \id-\im \sinh\tau {\bf u}), \Gamma\Gamma\Gamma\Gamma (\cosh\tau \id-\im \sinh\tau {\bf u})\rangle =\cosh(2\tau) (\id\wedge {\rm Im}(\Omega) +{\bf u} \wedge {\rm Re}(\Omega)) \\ \nonumber
-\im \sinh(2\tau) ( \id \wedge {\rm Re}(\Omega) - {\bf u}\wedge {\rm Im}(\Omega)) - \frac{1}{2} ( \id\wedge{\bf u} -\omega)\wedge ( \id\wedge{\bf u} -\omega).
\ee
We note that the blocks that appear here, namely $\id\wedge {\rm Im}(\Omega) +{\bf u} \wedge {\rm Re}(\Omega)$ and $\id \wedge {\rm Re}(\Omega) - {\bf u}\wedge {\rm Im}(\Omega)$ are the imaginary and real parts of the holomorphic 4-form (\ref{1u-omega}). Both are thus ${\rm SU}(4)$ invariant. The last term is the product of the two copies of the K\"ahler 2-form, which is again ${\rm SU}(4)$ invariant. This shows how the 4-form $B_4(\psi,\psi)$ is ${\rm SU}(4)$ invariant.

\section{Spin(4,4) and Split Octonions}
\label{sec:spin-44}

\subsection{Real model}

The link to split octonions arises if we consider Majorana-Weyl spinors. These are easiest to describe in the real model that starts by selecting a paracomplex structure on $\R^{4,4}$. However, Majorana-Weyl real spinors do not capture all possible spinor types that arise, as we have witnessed in the previous section. For this reason, it is better to develop everything in a complex model from the beginning, and then impose the Majorana condition if needed. Nevertheless, we start by describing the simpler real model, and then switch to the complex description. 

The real model arises by selecting a paracomplex structure on $\R^{4,4}$. Let $\R^4$ be one of the arising totally null subspaces, and let $u_I, I=1,\ldots,4$ be the null coordinates and $du_I$ the basic one-forms. It is more convenient to use the notation $du_I= e_I$. We introduce four pairs of creation/annihilation operators $a_I,a_I^\dagger$. We define the $\Gamma$-operators as follows
\begin{equation}
    \begin{split}
        \Gamma_0&=a_4+a_4^{\dagger},\\
        \Gamma_1&=a_1+a_1^{\dagger},\\
        \Gamma_2&=a_2+a_2^{\dagger},\\
        \Gamma_3&=a_3+a_3^{\dagger},\\
    \end{split}
\qquad    
    \begin{split}
        \Gamma_4&=a_4-a_4^{\dagger},\\
        \Gamma_5&=a_1-a_1^{\dagger},\\
        \Gamma_6&=a_2-a_2^{\dagger},\\
        \Gamma_7&=a_3-a_3^{\dagger}.\\
    \end{split}
\end{equation}

The Majorana-Weyl spinors are even and odd polyforms in $\Lambda(\R^4)$ with real coefficients. The basic polyforms $e_{IJ\ldots}:= e_I \wedge e_J \wedge \ldots $ are all null with respect to the invariant inner product that will be written below. For this reason, to make link with the split octonions in the usual non-null basis we introduce a non-null basis in $\Lambda(\R^4)$. We write
\begin{equation}
    \begin{split}
  S^+\ni  \psi^+= \alpha_1(e_{41}-e_{23}) + \alpha_2(e_{42}-e_{31})+\alpha_3(e_{43}-e_{12})+\alpha_4(1-e_{4123}) \\
   +\alpha_5(e_{41}+e_{23}) + \alpha_6(e_{42}+e_{31})+\alpha_7(e_{43}+e_{12})+\alpha_0(1+e_{4123}),
\\
 S^- \ni  \psi^-= \beta_1(e_{1}-e_{423}) + \beta_2(e_{2}-e_{431})+\beta_3(e_{3}-e_{412})+\beta_4(e_4-e_{123}) \\
   +\beta_5(e_{1}+e_{423}) + \beta_6(e_{2}+e_{431})+\beta_7(e_{3}+e_{412})+\beta_0(e_4+e_{123}).       
    \end{split}
\end{equation}
The invariant inner product is a pairing $\langle S^+, S^+\rangle, \langle S^-, S^-\rangle$. A simple computation gives
\be
\langle \psi^+,\psi^+\rangle = 2( (\alpha_0)^2 + (\alpha_1)^2+(\alpha_2)^2+(\alpha_3)^2 - (\alpha_4)^2- (\alpha_5)^2-(\alpha_6)^2-(\alpha_7)^2), \\ \nonumber
\langle \psi^-,\psi^-\rangle = 2( (\beta_0)^2 + (\beta_1)^2+(\beta_2)^2+(\beta_3)^2 - (\beta_4)^2- (\beta_5)^2-(\beta_6)^2-(\beta_7)^2).
\ee

We now form a Dirac spinor, which is a 16-component column
\be\label{ab-column}
\Psi = \left( \begin{array}{c} \psi^+ \\ \psi^-\end{array}\right) = \left( \begin{array}{c} \alpha_0 \\ \alpha_1 \\ \vdots \\ \alpha_7 \\ \beta_0 \\ \beta_1 \\ \vdots \\ \beta_7  \end{array}\right),
\ee
where we put the 8 $\alpha$ components of $\psi^+$ on top and 8 $\beta$ components of $\psi^-$ at the bottom of the column. The order in which the components appear is $0,1,\ldots,7$. The $\Gamma$-operators become the following $\Gamma$-matrices in this basis
\begin{equation}\label{Gamma-matr-spin44}
    \Gamma_0=
    \begin{pmatrix}
    0&\id\\
    \id&0
    \end{pmatrix}
    , \quad
    \Gamma_a=
    \begin{pmatrix}
    0&-\tilde{E}_a\\
    \tilde{E}_a&0
    \end{pmatrix}
    \ \text{for} \ a\in \{1,\ldots,7\},
\end{equation}
where
\begin{equation}\label{E-matr-spin44}
    \begin{split} 
        \tilde{E}_1&=-E_{01}-E_{23}-E_{45}+E_{67}\\
        \tilde{E}_2&=-E_{02}-E_{31}-E_{46}+E_{75}\\
        \tilde{E}_3&=-E_{03}-E_{12}-E_{47}+E_{56}\\
        \tilde{E}_4&=S_{04}-S_{15}-S_{26}-S_{37}\\
        \tilde{E}_5&=S_{05}+S_{14}-S_{27}+S_{36}\\
        \tilde{E}_6&=S_{06}+S_{17}+S_{24}-S_{35}\\
        \tilde{E}_7&=S_{07}-S_{16}+S_{25}+S_{34}.
    \end{split}
\end{equation}

\subsection{Split octonions}

Split octonions $\Oc'$ form a non-associative normed composition algebra. It is not a division algebra because there are null elements. A split octonion is an object 
\be
\tilde{q} = \tilde{q}_0 \id + \sum_{a=1}^7 \tilde{q}_a {\bf\tilde{e}}^a.
\ee
The unit imaginary octonions ${\bf\tilde{e}}^a$ anti-commute and satisfy
\be
({\bf\tilde{e}}^1)^2=({\bf\tilde{e}}^2)^2=({\bf\tilde{e}}^3)^2=-\id, \quad
({\bf\tilde{e}}^4)^2=({\bf\tilde{e}}^5)^2=({\bf\tilde{e}}^6)^2=({\bf\tilde{e}}^7)^2=\id.
\ee
Thus, the split octonions $\id, {\bf\tilde{e}}^1, {\bf\tilde{e}}^2, {\bf\tilde{e}}^3$ generate a copy of $\Hq\subset\Oc'$. The octonion pairing is given by
\be
(\tilde{q},\tilde{q}) = \tilde{q} \overline{\tilde{q}} = (\tilde{q}_0)^2 + (\tilde{q}_1)^2+ (\tilde{q}_2)^2+ (\tilde{q}_3)^2- (\tilde{q}_4)^2-(\tilde{q}_5)^2-(\tilde{q}_6)^2-(\tilde{q}_7)^2,
\ee
where the conjugation denoted by overbar changes the signs of all the imaginary generators. 

The product rules are most efficiently encoded into the following 3-form on ${\rm Im}(\Oc')=\R^7$
\begin{equation}\label{tilde-C}
    \tilde{C}={\bf\tilde{e}}^{123}-{\bf\tilde{e}}^{1}({\bf\tilde{e}}^{45}-{\bf\tilde{e}}^{67})-{\bf\tilde{e}}^2({\bf\tilde{e}}^{46}-{\bf\tilde{e}}^{75})-{\bf\tilde{e}}^3({\bf\tilde{e}}^{47}-{\bf\tilde{e}}^{56}).
\end{equation}
This encodes the vector product via 
\be
(u\times v, w) = w \lrcorner v\lrcorner u \lrcorner \tilde{C}.
\ee
Here $u\lrcorner$ is the operator of insertion of a vector field $u$ into a differential form. For example ${\bf\tilde{e}}^1 \times {\bf\tilde{e}}^2={\bf\tilde{e}}^3$, but ${\bf\tilde{e}}^1 \times {\bf\tilde{e}}^6=-{\bf\tilde{e}}^7$ because the octonion pairing is negative-definite on directions $4,5,6,7$. 

 \subsection{Octonionic model for ${\rm Cl}(4,4)$}

One can encode the operators of left multiplication by a unit octonion into $8 \times 8$ matrices. Indeed, we encode an octonion into an 8-component column
\be
\tilde{q} \to \left(\begin{array}{c} \tilde{q}_0 \\ \tilde{q}_1 \\ \vdots \\ \tilde{q}_7\end{array}\right).
\ee
It is then a straightforward computation to see that the operators of left multiplication by a unit octonion precisely match the matrices in (\ref{E-matr-spin44}) $L_{\bf{\tilde{e}}^a} = \tilde{E}_a$. This shows that a general linear combination of the $\Gamma$-matrices (\ref{Gamma-matr-spin44}) is a matrix of the form (\ref{X}) with $r=0$
\be
\tilde{q}_0 \Gamma_0 + \sum_{a=1}^7 \tilde{q}_a \Gamma_a = \left( \begin{array}{cc} 0 & L_{\overline{\tilde{q}}} \\ L_{\tilde{q}} & 0 \end{array}\right),
\ee
where $L_{\tilde{q}}$ is the operator of left multiplication by a split octonion $\tilde{q}\in \Oc'$. This shows how the model (\ref{X}) described in the Introduction, with $q\in \Oc'$ and $r=0$, arises from the creation/annihilation operator model, in the version of it which uses real polyforms. 

\subsection{The complex model}

We now develop the complex model of ${\rm Cl}(4,4)$. The starting point is a complex structure on $\R^{4,4}$, so that $\R^{4,4}$ gets identified with $\C^4$. Let $z_I, I=1,\ldots, 4$ be the corresponding complex (null) coordinates, and $e^I = dz_I$ the basic one-forms. The metric on $\R^{4,4}$ becomes the following indefinite Hermitian metric
\be
|dz_3|^2 + |dz_4|^2 - |dz_1|^2 - |dz_2|^2.
\ee
A general Dirac spinor is a polyform in $\Lambda(\C^4)$, with complex coefficients. The $\Gamma$-operators that square to plus the identity are given by
\be
\Gamma_0 = a_4+a_4^\dagger, \qquad \Gamma_3 = \im (a_4-a_4^\dagger), \\
\nonumber
\Gamma_2= a_3+a_3^\dagger, \qquad \Gamma_1 = \im (a_3-a_3^\dagger).
\ee
Note that these generate a copy of ${\rm Cl}(4)$, and act only on the $e_3, e_4$ polyform directions. The $\Gamma$-operators that square to minus the identity are
\be
\Gamma_4 = \im (a_2+a_2^\dagger), \qquad \Gamma_7 = a_2-a_2^\dagger, \\
\nonumber
\Gamma_6 = \im (a_1+a_1^\dagger), \qquad \Gamma_5 = a_1-a_1^\dagger.
\ee
The link to split octonions arises if we consider Majorana-Weyl spinors, so we must understand the reality conditions first.

\subsection{Reality conditions}

There are two anti-linear operators, the product of all real $\Gamma$-operators followed by the complex conjugation, and the product of the imaginary ones followed by the complex conjugation. Both square to plus the identity, and so give a possible reality condition. They only differ in their action (by a sign) on odd polyforms, and agree on even polyforms. It turns out to be better to use 
\be
R'= \Gamma_3\Gamma_1\Gamma_4\Gamma_6 \ast
\ee
as the reality condition. A simple calculation shows that the following polyforms parametrised by $\C^4$ are real
\be
\psi_- = u_1 e^1 - u_1^* e^{423} + u_2 e^2 - u_2^* e^{431} + u_3 e^3 + u_3^* e^{412} + i u_4^* e^4 + \im u_4 e^{123}, \\ \nonumber
\psi_+ = u_1 e^{41} - u_1^* e^{23} + u_2 e^{42} - u_2^* e^{31} + u_3 e^{43} + u_3^* e^{12} + \im u_4^* + \im u_4 e^{4123}.
\ee
This should be compared to (\ref{real-polyforms}). Only some signs are different as compared to the ${\rm Cl}(8)$ situation. We parametrise the even polyforms by the real and imaginary parts of $u_I$ 
\be
u_1= \alpha_5+\im\alpha_6, \quad u_2 =\alpha_7 +\im \alpha_4, \quad
u_3= \alpha_2 + \im \alpha_1, \quad u_4= \alpha_3 +\im \alpha_0.
\ee
We get the following real parametrisation of even and odd polyforms
\be
\psi^+ = \alpha_5( e^{41}-e^{23}) + \alpha_7( e^{42}-e^{31}) +\alpha_2( e^{43}+e^{12}) + \alpha_0 (1- e^{4123}) \\
\nonumber
+ \im \alpha_6( e^{41}+e^{23}) + \im \alpha_4( e^{42}+e^{31}) +\im \alpha_1( e^{43}-e^{12}) + \im \alpha_3 (1+ e^{4123}) ,
\\ 
\psi^- = \beta_5( e^{1}-e^{423}) + \beta_7( e^{2}-e^{431}) +\beta_2( e^{3}+e^{412}) + \beta_0 (e^4- e^{123}) \\
\nonumber
+ \im \beta_6( e^{1}+e^{423}) + \im \beta_4( e^{2}+e^{431}) +\im \beta_1( e^{3}-e^{412}) + \im \beta_3 (e^4+ e^{123}).
\ee
The spinor norms are then
\be
\frac{1}{2} \langle \psi_+,\psi_+\rangle = (\alpha_0)^2 + (\alpha_1)^2 + (\alpha_2)^2+(\alpha_3)^2 - (\alpha_4)^2-(\alpha_5)^2 - (\alpha_6)^2 - (\alpha_7)^2, \\ \nonumber
\frac{1}{2} \langle \psi_-,\psi_-\rangle = (\beta_0)^2 + (\beta_1)^2 + (\beta_2)^2+(\beta_3)^2 - (\beta_4)^2-(\beta_5)^2 - (\beta_6)^2 - (\beta_7)^2.
\ee
We now place the components $\alpha,\beta$ into a 16-component column (\ref{ab-column}), and work out the matrix representation of the $\Gamma$-operators. We get precisely the matrices of the form (\ref{Gamma-matr-spin44}) with (\ref{E-matr-spin44}), which also justifies the choices for the signs of the $\Gamma$-operators. So, we again reproduce the model (\ref{X}) explained in the Introduction, with $q\in \Oc'$. 

\subsection{Pure spinors}

In contrast to the ${\rm Spin}(8)$ case, we now have several different types of pure spinors. We describe them all on the basis of the complex model.

The complex model was obtained by choosing a complex structure on $\R^{4,4}$, and so the canonical pure spinors that this model comes with, namely $1, e^{4123}\in S_+$ give back this complex structure. To see this, we translate the polyforms into split octonions. As in the case of ${\rm Spin}(8)$ we need to complexify the octonions to see (at least certain types of) the pure spinors. We have
\be
{\bf\tilde e}^3 = \im (1+ e^{4123}), \qquad \id = (1- e^{4123}),
\ee
and so 
\be\label{pure44-compl}
1= \frac{1}{2}(\id -\im {\bf\tilde e}^3), \qquad e^{4123} = -\frac{1}{2}(\id +\im {\bf\tilde e}^3).
\ee

We now need the following results
\be
( (\id +\im {\bf\tilde e}^3), L_x (\id -\im {\bf\tilde e}^3)) = 2\im {\bf\tilde e}^3, \\ \nonumber
( (\id +\im {\bf\tilde e}^3), L_x L_y (\id -\im {\bf\tilde e}^3)) =2\im ( {\bf\tilde e}^{12} + {\bf\tilde e}^{74}+{\bf\tilde e}^{56}).
\ee
This means that
\be
\langle (\id +\im {\bf\tilde e}^3), \Gamma\Gamma (\id -\im {\bf\tilde e}^3)\rangle = 2\im ( \id\wedge {\bf\tilde e}^3 - ( {\bf\tilde e}^{12} + {\bf\tilde e}^{74}+{\bf\tilde e}^{56}).
\ee
This is a $(1,1)$ form of the complex structure that this (complex conjugate) pair of pure spinors defines. To see what this complex structure is, let us consider the right multiplication by ${\bf\tilde e}^3$. The $-\im$ eigenvectors of $R_{{\bf\tilde e}^3}$ are
\be
z_3:= \id +\im {\bf\tilde e}^3, \qquad z_4:= {\bf\tilde e}^1 -\im {\bf\tilde e}^2, \qquad z_1:= {\bf\tilde e}^7 +\im {\bf\tilde e}^4, \qquad z_2:= {\bf\tilde e}^5 +\im {\bf\tilde e}^6.
\ee
We then have
\be
\langle (\id +\im {\bf\tilde e}^3), \Gamma\Gamma (\id -\im {\bf\tilde e}^3)\rangle 
= z_3^*\wedge z_3 + z_4^*\wedge z_4 - z_1^*\wedge z_1-z_2^*\wedge z_2,
\ee 
and so indeed $R_{{\bf\tilde e}^3}$ is the complex structure that corresponds to the pure spinors $1,e^{4123}$. 

Let us also state the result for the stabiliser of the spinor $\id -\im {\bf\tilde e}^3$. The general Lie algebra element on $S_+$ is
\be
A_{\mathfrak{spin}(4,4)} = w^a \tilde{E}_a - w^{ab} \tilde{E}_a \tilde{E}_b.
\ee
The stabiliser of $\id -\im {\bf\tilde e}^3$ is the subalgebra determined by the following equations
\be
w^3=w^{12}+w^{56}+w^{74}, \\ \nonumber
w^1 = w^{23}, \quad w^{45}=w^{67}, \quad w^2=w^{31}, \quad w^{46}=w^{75}, \\ \nonumber
w^4 = w^{73}, \quad w^{51}=w^{26}, \quad w^7=w^{34}, \quad w^{16}=w^{25}, \\ \nonumber
w^5 = w^{36}, \quad w^{14}=w^{27}, \quad w^6=w^{53}, \quad w^{71}=w^{24}.
\ee
The stabiliser is thus $28-13=15$ dimensional. Given that it preserves a complex structure in $\R^{4,4}$, the stabiliser coincides with $\mathfrak{su}(2,2)$. 

A different type of pure spinors that is not difficult to describe corresponds to a paracomplex structure on $\R^{4,4}$. Consider the null split octonions
\be
\frac{1}{2}(\id+{\bf\tilde e}^4)= 1-e^{4123}+\im(e^{42}+e^{31}), \qquad \frac{1}{2}(\id-{\bf\tilde e}^4)= 1-e^{4123}-\im(e^{42}+e^{31}),
\ee
where we also indicated the corresponding polyforms. We have
\be
( (\id - {\bf\tilde e}^4), L_x (\id + {\bf\tilde e}^4)) = 2 {\bf\tilde e}^4, \\ \nonumber
( (\id - {\bf\tilde e}^4), L_x L_y (\id + {\bf\tilde e}^4)) =-2 ( {\bf\tilde e}^{15} + {\bf\tilde e}^{26}+{\bf\tilde e}^{37}).
\ee
Therefore
\be
\langle (\id - {\bf\tilde e}^4), \Gamma\Gamma (\id + {\bf\tilde e}^4)\rangle = 2( \id\wedge {\bf\tilde e}^4 +  {\bf\tilde e}^{15} + {\bf\tilde e}^{26}+{\bf\tilde e}^{37}).
\ee
We thus see that the pair of pure spinors $\id - {\bf\tilde e}^4,\id + {\bf\tilde e}^4$ defines the paracomplex structure whose real null eigenvectors are
\be
\id+ {\bf\tilde e}^4, \qquad {\bf\tilde e}^1+{\bf\tilde e}^5, \qquad {\bf\tilde e}^2+{\bf\tilde e}^6, \qquad {\bf\tilde e}^3+{\bf\tilde e}^7.
\ee
As an operator on $\Oc'$ this paracomplex structure is described by $R_{{\bf\tilde e}^4}$. 

Let us state the stabiliser in this case. The stabiliser of $\id+{\bf\tilde e}^4$ is given by the following equations
\be\label{stab-null-oct}
w^4+w^{15}+w^{26}+w^{37}=0, \\ \nonumber
w^1-w^{14}+w^5+w^{45} =0, \quad w^{23}+w^{27}-w^{36}+w^{67}=0, \\ \nonumber
w^2-w^{24}+w^6+w^{46}=0, \quad w^{13}+w^{17}-w^{35}+w^{57}=0, \\ \nonumber
w^3-w^{34}+w^7 +w^{47}=0, \quad w^{12}+w^{16}-w^{25}+w^{56}=0,
\ee
and so is $28-7=21$ dimensional. It can be understood geometrically as ${\rm SL}(4,\R)$ mixing up the four real null directions of this pure spinor, semi-direct product with a copy of nilpotent ${\rm SO}(4)$. If we in addition impose the condition that $\id-{\bf\tilde e}^4$ is fixed, we get six new equations 
\be
w^1+w^{14}-w^5+w^{45} =0, \quad w^{23}-w^{27}+w^{36}+w^{67}=0, \\ \nonumber
w^2+w^{24}-w^6+w^{46}=0, \quad w^{13}-w^{17}+w^{35}+w^{57}=0, \\ \nonumber
w^3+w^{34}-w^7 +w^{47}=0, \quad w^{12}-w^{16}+w^{25}+w^{56}=0,
\ee
which together with the previous set gives a subalgebra of dimension $28-7-6=15$. It is clear that the stabiliser is ${\rm SL}(4,\R)$ that mixes the four real null directions of the null subspaces of both pure spinors. 

Let us now consider the spinors
\be
\frac{1}{2} (\id- \im {\bf\tilde e}^3) - \frac{1}{2} ({\bf\tilde e}^4+\im {\bf\tilde e}^7) = 1 + \im e^{42},\qquad 
\frac{1}{2} (\id+ \im {\bf\tilde e}^3) + \frac{1}{2} ({\bf\tilde e}^4-\im {\bf\tilde e}^7) = -e^{4123} + \im e^{31}.
\ee
Both are null spinors, with a non-vanishing inner product between them. The first of them is annihilated by
\be
\Gamma_1-\im \Gamma_2, \quad \Gamma_5+\im \Gamma_6, \quad \Gamma_0+\Gamma_4, \quad \Gamma_3+\Gamma_7,
\ee
and the second by the complement of these four vectors. So, they are a pair of pure spinors with null subspace spanned by two real and two complex vectors. Note that we can rewrite these pure spinors as
\be\label{pure44-index2}
\frac{1}{2} (\id-  {\bf\tilde e}^4) - \frac{\im}{2} ({\bf\tilde e}^3+ {\bf\tilde e}^7), \qquad 
\frac{1}{2} (\id+ {\bf\tilde e}^4) + \frac{\im}{2} ({\bf\tilde e}^3- {\bf\tilde e}^7).
\ee
Both are of the form $\alpha+\im\beta$ where $\alpha,\beta$ are real pure spinors with $(\alpha,\beta)=0$. 

We want to see how the structure of the mixed type gets produced by these spinors. We have
\be
( (\id- \im {\bf\tilde e}^3 - {\bf\tilde e}^4-\im {\bf\tilde e}^7), L_x (\id+\im {\bf\tilde e}^3 + {\bf\tilde e}^4-\im {\bf\tilde e}^7)) = 4 {\bf\tilde e}^4, \\ \nonumber
( (\id- \im {\bf\tilde e}^3 - {\bf\tilde e}^4-\im {\bf\tilde e}^7), L_x L_y (\id+\im {\bf\tilde e}^3 + {\bf\tilde e}^4-\im {\bf\tilde e}^7)) =-4\im ( {\bf\tilde e}^{12} + {\bf\tilde e}^{56}-\im{\bf\tilde e}^{37}).
\ee
This means that we have
\be
\langle (\id- \im {\bf\tilde e}^3 - {\bf\tilde e}^4-\im {\bf\tilde e}^7), \Gamma\Gamma (\id+\im {\bf\tilde e}^3 + {\bf\tilde e}^4-\im {\bf\tilde e}^7)\rangle = 4( \id\wedge {\bf\tilde e}^4 +{\bf\tilde e}^{37}+ \im {\bf\tilde e}^{12} + \im {\bf\tilde e}^{56}).
\ee
This is a complex tensor of a mixed type. Raising one of the indices gives a complex endomorphism of $\R^{4,4}$ that squares to minus the identity, and is a sum of a paracomplex structure in the directions $\id, 4,3,7$ and the imaginary unit times the complex structure in the directions $1,2,5,6$. 

Let us determine the stabilisers in this case. The stabiliser of $(\id- \im {\bf\tilde e}^3 - {\bf\tilde e}^4-\im {\bf\tilde e}^7)$ is given by the following set of equations
\be
w^{16}-w^{25}=0, \quad w^{12}+w^{56}=0, \quad w^{15}+w^{26}=0, \quad w^{37}+w^4=0, \\ \nonumber
w^1+w^{14}=0, \quad w^{27}-w^{23}=0, \quad w^2+w^{24}=0, \quad w^{35}+w^{57}=0, \\ \nonumber
w^5-w^{45}=0, \quad w^{36}+w^{67}=0,\quad w^6-w^{46}=0, \quad w^{13}-w^{17}=0, \\ \nonumber
w^3+w^{34}+w^{47}-w^7=0.
\ee
The dimension of the stabiliser is thus $28-13=15$. Demand that also the complementary spinor $(\id+ \im {\bf\tilde e}^3 + {\bf\tilde e}^4-\im {\bf\tilde e}^7)$ is stabilised reproduces the equations in the first line, while gives for the other lines
\be
w^1-w^{14}=0, \quad w^{27}+w^{23}=0, \quad w^2-w^{24}=0, \quad w^{35}-w^{57}=0, \\ \nonumber
w^5+w^{45}=0, \quad w^{36}-w^{67}=0,\quad w^6+w^{46}=0, \quad w^{13}+w^{17}=0, \\ \nonumber
w^3-w^{34}+w^{47}+w^7=0.
\ee
Thus, together the equations imply
\be
w^{16}-w^{25}=0, \quad w^{12}+w^{56}=0, \quad w^{15}+w^{26}=0, \quad w^4=w^{73}, \quad 
w^3=w^{74}, \quad w^7=w^{34},
\ee
as well as
\be
w^{1,2,5,6}=0, \quad w^{41}=w^{42}=w^{45}=w^{46}=0, \quad w^{27}=w^{23}=w^{35}=w^{57}=w^{36}=w^{67}=w^{13}=w^{17}=0.
\ee
The stabiliser of both complementary pure spinors is thus 6-dimensional. The geometric interpretation of the stabiliser is that it consists of ${\rm SU}(1,1)$ mixing up the two complex null directions ${\bf\tilde e}^1-\im {\bf\tilde e}^2, {\bf\tilde e}^5+\im {\bf\tilde e}^6$, as well as ${\rm SL}(2,\R)$ mixing up the two real null directions $\id+ {\bf\tilde e}^4, {\bf\tilde e}^3+{\bf\tilde e}^7$.

We have thus described the three different types of pure spinors of ${\rm Spin}(4,4)$. One gives a complex structure, one paracomplex, and the third type gives a structure of the mixed type. Of these only the pure spinor giving the paracomplex structure is real. 

\subsection{Majorana spinors}

The Majorana spinors are simply the split octonions. There are three possible types of such spinors. Non-null spacelike or timelike, and null. We have determined the stabiliser of a null octonion in (\ref{stab-null-oct}). As we have seen in the previous subsection, a null octonion is a pure spinor. The stabilisers of spacelike or timelike octonions are also 21 dimensional and in both cases are given by ${\rm Spin}(4,3)$. 

For future usage we compute explicitly the stabiliser subalgebra of $\psi=\id$. It is given by the following relations
\be\label{spin7}
w^1= w^{23}-w^{45}+w^{67}, \quad w^2 = -w^{13}-w^{46}-w^{57}, \quad w^3= w^{12}-w^{47}+w^{56}, \\ \nonumber 
w^4= -w^{15}-w^{26}-w^{37}, \quad w^5=w^{14}-w^{27}+w^{36}, \quad w^6 = w^{17}+w^{25}-w^{35}, \\ \nonumber
w^7 = -w^{16}+w^{25}+w^{34}.
\ee

\subsection{General spinors}

We now enter into a less familiar territory, as there seems to be no known classification of the orbits of the real ${\rm Spin}(4,4)$ on the complex Weyl spinors, apart from already considered case of pure and Majorana spinors. This is in contrast to the case of ${\rm Spin}(8)$, where there is only one possible type of general spinors, with the stabiliser ${\rm SU}(4)$.

To classify general ${\rm Spin}(4,4)$ spinors we use the same idea that worked in the ${\rm Spin}(8)$ case. We consider a general complex spinor, which is a complexified split octonion $\psi=\alpha+\im \beta$. The relations (\ref{psi-alpha-beta}) are still valid. The only novelty now is that the norm squared does not need to be positive. We assume that the spinor is not null (because if it is null it is pure). We again rescale the spinor to make it unit, so that $|\alpha|^2-|\beta|^2=1$ and $(\alpha,\beta)=0$. This gives $\langle R(\psi),\psi\rangle = 1+2|\beta|^2$. The novelty now is that this quantity does not need to be greater than one. 

There are five cases to consider. First, when $|\beta|^2>0$ both $\alpha,\beta$ have positive norm and can be chosen to be multiples of $\id,{\bf\tilde{e}}^3$. We get the unit spinor
\be
\psi= \cosh\tau \id + \im \sinh\tau {\bf\tilde{e}}^3.
\ee
It is clear that the analysis in the case of ${\rm Spin}(8)$ is unchanged, and this spinor defines a pure spinor that is a multiple of $\id+\im {\bf\tilde{e}}^3$, whose stabiliser is ${\rm SU}(2,2)$. Thus, the general spinor of this type still defines a complex structure, and its stabiliser is ${\rm SU}(2,2)$.

Another case is when $-1<|\beta|^2<0$. This means that $|\alpha|^2>0$, and we can still choose $\alpha$ to be a multiple of $\id$. This leads us to consider the unit spinor
\be
\psi= \cos\theta \id + \im \sin\theta\, {\bf\tilde{e}}^4.
\ee
We have $\langle R(\psi),\psi\rangle = \cos^2\theta-\sin^2\theta= \cos(2\theta)$. Denoting $\lambda=\cos(2\theta)$ we can form
\be
\tilde{\psi}= \frac{\lambda\psi - R(\psi)}{\im\sqrt{1-\lambda^2}} = \im\sin\theta \, \id + \cos\theta \, {\bf\tilde{e}}^4.
\ee 
This is again a spinor of norm minus one, and so 
\be
\psi\pm\tilde{\psi} = (\cos\theta+\im \sin\theta) ( \id \pm {\bf\tilde{e}}^4)
\ee
are both null and thus pure spinors. They are (complex) multiples of the spinors $( \id \pm {\bf\tilde{e}}^4)$ that we already encountered before. This pair of pure spinor defines a paracomplex structure on $\R^{4,4}$. Thus, the general spinors of this type defines two pure spinors of the type $( \id \pm {\bf\tilde{e}}^4)$, and thus defines a paracomplex structure on $\R^{4,4}$. The stabiliser of a general spinor of this type is thus ${\rm SL}(4,\R)$. 

Yet another case is $|\beta|^2<-1$. This means that both $\alpha,\beta$ have negative norms. For example we can choose $\alpha\sim{\bf\tilde{e}}^4, \beta\sim {\bf\tilde{e}}^7$. This leads us to consider
the unit spinor
\be
\psi = \sinh\tau \, {\bf\tilde{e}}^4+\im \cosh\tau \,{\bf\tilde{e}}^7.
\ee
Then $\lambda= \langle R(\psi),\psi\rangle = -\cosh(2\tau)$ and 
\be
\tilde{\psi} = \frac{\lambda\psi - R(\psi)}{\sqrt{\lambda^2-1}} = -\cosh\tau \, {\bf\tilde{e}}^4-\im \sinh\tau \,{\bf\tilde{e}}^7.
\ee
This is again a spinor of norm minus one, and the spinor
\be
\psi+\tilde{\psi} = (\sinh\tau-\cosh\tau) ( {\bf\tilde{e}}^4-\im {\bf\tilde{e}}^7)
\ee
is pure. This spinor defines a complex structure on $\R^{4,4}$ with the K\"ahler form
\be
\langle ( {\bf\tilde{e}}^4-\im {\bf\tilde{e}}^7),\Gamma\Gamma ( {\bf\tilde{e}}^4+\im {\bf\tilde{e}}^7)\rangle = 
2\im (\id \wedge {\bf\tilde{e}}^3 + {\bf\tilde{e}}^{12}+{\bf\tilde{e}}^{47}+{\bf\tilde{e}}^{56}).
\ee
Thus, the general spinor of this type still defines a complex structure, and its stabiliser is ${\rm SU}(2,2)$. 

There are also two cases when a null split octonion arises. When $\beta$ is null, the octonion $\alpha$ is unit, and we are led to consider the spinor
\be
\psi = \id + \im ( {\bf\tilde{e}}^3 + {\bf\tilde{e}}^4).
\ee
This is complex spinor whose real part is an identity octonion, and the imaginary part is a null octonion. Then $\lambda=\langle R(\psi),\psi\rangle = 1$ and the construction of $\tilde{\psi}$ is no longer applicable. It is clear that the geometry arising in this case knows both about the geometry related to $\id$, and that of the real pure spinor ${\bf\tilde{e}}^3 + {\bf\tilde{e}}^4$. The identity octonion defines the ${\rm Spin}(7)$ invariant 4-form $-\id\wedge \tilde{C}+ {}^*\tilde{C}$ on $\R^{4,4}$. The pure spinor ${\bf\tilde{e}}^3 + {\bf\tilde{e}}^4$ defines its null subspace that can be seen to be spanned by
\be\label{34-MTN}
\id-  {\bf\tilde{e}}^7 , \quad {\bf\tilde{e}}^3+{\bf\tilde{e}}^4, \quad {\bf\tilde{e}}^2 -  {\bf\tilde{e}}^5, \quad 
{\bf\tilde{e}}^1+ {\bf\tilde{e}}^6.
\ee
To understand the geometry arising better we compute
\be\label{2form-new-orbit}
\langle (\id + \im {\bf\tilde{e}}^3 + \im {\bf\tilde{e}}^4), \Gamma\Gamma (\id - \im {\bf\tilde{e}}^3 - \im {\bf\tilde{e}}^4)\rangle = 2\im ( (\id+  {\bf\tilde{e}}^7 )\wedge ({\bf\tilde{e}}^3-{\bf\tilde{e}}^4) 
+ ( {\bf\tilde{e}}^2 +  {\bf\tilde{e}}^5)\wedge (  {\bf\tilde{e}}^1- {\bf\tilde{e}}^6) ).
\ee
The 2-form that arises is thus a sum of two decomposable pieces, each built entirely from null vectors that are complementary to those in (\ref{34-MTN}).

It is also interesting to compute the stabiliser of $\id + \im {\bf\tilde{e}}^3 + \im {\bf\tilde{e}}^4$. It is clear that it is the intersection of the stabilisers of $\id$ and $ {\bf\tilde{e}}^3 +  {\bf\tilde{e}}^4$. Both are 21-dimensional. An explicit calculation shows that this intersection is given by (\ref{spin7}) supplemented by the following relations
\be
 w^{16}=w^{25}, \quad 
w^{12}+w^{15}+w^{26}+w^{56}=0, \quad w^{23}-w^{24}-w^{35}+w^{45}=0, \quad -w^{13}+w^{14}-w^{36}+w^{46}=0.
\ee
It is thus $21- 4= 17$ dimensional.

\section{Discussion}

The Clifford algebras ${\rm Cl}(3), {\rm Cl}(5), {\rm Cl}(9)$ and ${\rm Cl}(2,1), {\rm Cl}(3,2), {\rm Cl}(5,4)$ admit a uniform description as generated by matrices of the form (\ref{X}), where $q$ takes values in $\C,\Hq,\Oc$ for the first list and the split composition algebras $\C',\Hq',\Oc'$ for the second list. It is also easy to generate related models for Clifford algebras in one dimension up and one (or more) dimensions down. To go one dimension up one uses the tensor product construction and generates off-diagonal $2\times 2$ matrices where the off-diagonal block entries are the matrices (\ref{X}) as in (\ref{gamma1-5}). The additional $\Gamma$-matrix is then the matrix (\ref{gamma-last}). This again gives a uniform description of  ${\rm Cl}(3,1), {\rm Cl}(5,1), {\rm Cl}(9,1)$ and ${\rm Cl}(2,2), {\rm Cl}(3,3), {\rm Cl}(5,5)$. To go one dimension down one just sets $r=0$ in (\ref{X}), getting a description of
${\rm Cl}(2), {\rm Cl}(4), {\rm Cl}(8)$ and ${\rm Cl}(1,1), {\rm Cl}(2,2), {\rm Cl}(4,4)$. It is possible to go down in dimension even further. Thus, one can omit the $\Gamma$-matrix corresponding to the identity element in $\C,\Hq,\Oc,\C',\Hq',\Oc'$. This gives a model for ${\rm Cl}(1), {\rm Cl}(3), {\rm Cl}(7)$ and ${\rm Cl}(0,1), {\rm Cl}(1,2), {\rm Cl}(3,4)$. One can go down in dimension even further by selecting a unit imaginary element in $\C,\Hq,\Oc,\C',\Hq',\Oc'$ and then omitting the corresponding $\Gamma$-matrix. This gives a model for ${\rm Cl}(0), {\rm Cl}(2), {\rm Cl}(6)$ and ${\rm Cl}(0,0)$ for $\C,\Hq,\Oc,\C'$. Of these only the model for ${\rm Cl}(6)$ is of interest, and gives a useful and powerful octonionic description of ${\rm Spin}(6)$. In the case of $\Hq',\Oc'$ there is a choice of which imaginary element to omit, and one can get models for either ${\rm Cl}(0,2)$ or ${\rm Cl}(1,1)$ in the case of $\Hq'$, or ${\rm Cl}(2,4)$ and ${\rm Cl}(3,3)$ for $\Oc'$. This gives a split octonionic description of ${\rm Spin}(2,4)$ and ${\rm Spin}(3,3)$. Both are powerful and useful descriptions of these groups. Especially the first of these should be noted, as it is the conformal group of the Minkowski space. 

The described constructions are relatively well-known, see e.g. \cite{Harvey}, even though their split versions are rarely discussed in the literature. The novelty of this paper is that we have shown how the models based on $\C,\Hq,\Oc,\C',\Hq',\Oc'$ arise from the creation/annihilation operator description of the relevant Clifford algebras. Thus, we have seen that even and odd polyforms on $\C^2$ can be identified with either quaternions or split quaternions. The difference between these two cases arises only in the form of the anti-linear hat operator that acts on spinors and commutes with all $\Gamma$-matrices. In the case of ${\rm Cl}(4)$ the hat operator squares to minus the identity. It anti-commutes with the complex structure that acts on the components of spinors as the operator of multiplication by $-\im$. In the quaternionic description the complex structure that is the operator of multiplication by $-\im$ gets translated into the operator $R_{\bf k}$ of right multiplication by a unit imaginary quaternion. The hat operator gets translated into $-R_{\bf i}$. The two operators together generate the quaternionic structure. In the case of ${\rm Cl}(2,2)$ the hat operator squares to plus the identity. In the split quaternionic description we again have the complex structure that translates as $R_{\bf\tilde k}$. The hat operator translates into $R_{\bf\tilde i}$. Again this generates the quaternionic structure, but this time it is the split quaternions that appear. 

The difference in the cases of ${\rm Cl}(8)$ and ${\rm Cl}(4,4)$ is that even and odd polyforms on $\C^4$ {\bf satisfying a reality condition} are identified with octonions and split octonions. The anti-linear operator imposing the reality condition in both cases acts as the operator related to the Hodge star operator on $\Lambda(\C^4)$. The difference between ${\rm Cl}(8)$ and ${\rm Cl}(4,4)$ is in some signs arising. After the reality condition gets imposed, we can identify {\bf real} even and odd polyforms with either octonions or split octonions. Thus, Majorana-Weyl spinors of ${\rm Cl}(8)$ and ${\rm Cl}(4,4)$ are octonions and split octonions respectively. This gives a powerful description, because a general Weyl spinor can then be described as a complexified octonion or a split octonion. The derivation of the octonionic description from the one in terms of the creation/annihilation operators makes the description of pure spinors of ${\rm Cl}(8)$ and ${\rm Cl}(4,4)$ particularly clear. Pure spinors are null objects. In the case of ${\rm Cl}(8)$ they are null octonions and thus necessarily null complexified octonions. In the case of ${\rm Cl}(4,4)$ there are three types of pure spinors. There are now null split octonions, and they are pure spinors of real index four. The other two types of pure spinors are described by complexified octonions. There are two types that arise this way. In one type, the complexified null octonion has real and imaginary parts that are not null, see (\ref{pure44-compl}) for a representative of this orbit. This type of pure spinor defines a complex structure on $\R^{4,4}$. In the other case the real and imaginary parts are both null, see (\ref{pure44-index2}). This type of pure spinor defines the structure of a mixed type with two real and two complex null directions. We have also described the possible types of general spinors for both ${\rm Cl}(8)$ and ${\rm Cl}(4,4)$. In the case of ${\rm Cl}(4,4)$ there is an orbit that does not have a ${\rm Cl}(8)$ analog. It is the orbit when one of the two octonions in $\alpha+\im \beta$ is null. We have characterised its stabiliser and computed the arising in this case 2-form, see (\ref{2form-new-orbit}). 

The two constructions we described, namely the general description of the Clifford algebras using creation/annihilation operators, and the description of some of the Clifford algebras using ${\mathbb K}=\C,\Hq,\Oc,\C',\Hq',\Oc'$ allows to produce $\mathbb K$-based models for Clifford algebras other than those mentioned above. The idea is very simple. As discussed, even and odd polyforms on $\C^2$ and $\C^4$ can be identified with quaternions or split quaternions and with (complexified) octonions or split octonions. This means that polyforms on spaces containing the $\C^2$ and $\C^4$ factors can also be given a description based on $\Hq,\Hq',\Oc,\Oc'$. We have already seen how this works for ${\rm Cl}(6)$ in Section \ref{sec:spin6}. One can apply essentially the same construction to ${\rm Cl}(10)$ and generate its octonionic description. The resulting octonionic formalism has already been described in \cite{Bryant}. The semi-spinors of ${\rm Spin}(10)$ are then 2-component columns with complexified octonions as entries. If desired, one can go in dimension even higher and describe Clifford algebras such as ${\rm Cl}(12)$ and ${\rm Cl}(11,1)$ using octonions. Semi-spinors of both are 4-component columns with complexified octonions as entries. We hope to return to the octonionic description of the phenomenologically important cases of ${\rm Spin}(10)$ and ${\rm Spin}(11,1)$ in another publication. 

The starting point of a creation/annihilation operator model of ${\rm Cl}(r,s)$ is a mixed type structure on $\R^{r,s}$, see the accompanying paper \cite{BK}. This selects two complementary totally null subspaces $E^\pm$ of the complexification of $\R^{r,s}$. The spinors in the creation/annihilation operator description are then polyforms on one of the two totally null subspaces. In particular, even model comes with two "canonical" pure spinors whose annihilator subspaces are $E^\pm$. 

The models based on composition algebras $\mathbb K$ are different. The natural question is what is the geometric structure (if any) that a model of this type introduces. For concreteness, let us discuss this question for the case of ${\rm Spin}(8)$. As we have seen, the $\Oc$-based model identifies Majorana-Weyl spinors of ${\rm Spin}(8)$ with octonions. A Majorana-Weyl spinor of ${\rm Spin}(8)$ is never pure, and so there are no preferred pure spinors that arise from the model. The spinors that can be said to come with the model are the positive and negative helicity Majorana-Weyl spinors that correspond to the identity octonion. The stabiliser of one of them is ${\rm Spin}(7)$, the stabiliser of both of them is the group of automorphisms of the octonions ${\rm G}_2$. So, it could be said that the geometry on $\R^8$ that comes with the $\Oc$-based model of ${\rm Spin}(8)$ is the one that reduces the group ${\rm Spin}(8)$ to ${\rm G}_2$. Geometrically, a pair $\psi_\pm$ of Majorana-Weyl spinors of opposite helicity defines a vector in $\R^8$, via $\langle \psi_+, \Gamma\psi_- \rangle$. Also, each of the two Majorana-Weyl spinors defines the 4-form (\ref{4-form}), with the relative sign between the two terms in this form different for the two spinors. Inserting the vector $\id\in\Oc\sim \R^8$ into this 4-form one gets the 3-form $C\in \Lambda^3(\R^7)$, whose stabiliser is ${\rm G}_2$. This is the geometry defined by the two "canonical" Majorana-Weyl spinors $\id\in S_\pm$ that the $\Oc$-based model of ${\rm Spin}(8)$ comes with.

\section*{Acknowledgement} KK is grateful to F. Reese Harvey for correspondence.

\end{document}